\documentclass[sn-apa,iicol]{sn-jnl}

\usepackage{import}
\usepackage{subcaption}
\usepackage{bm}
\usepackage{amsmath}
\usepackage{mdframed}
\usepackage{algorithm}
\usepackage{algpseudocode}

\mdfdefinestyle{MyFrame}{%
    linecolor=black,
    outerlinewidth=4pt,
    innertopmargin=\baselineskip,
    innerbottommargin=\baselineskip,
    innerrightmargin=20pt,
    innerleftmargin=20pt,
    backgroundcolor=gray!5!white,
    roundcorner=20pt}

\algnewcommand{\Inputs}[1]{%
  \Statex \text{Inputs:} {\raggedright #1}
}

\algnewcommand{\Initialize}[1]{%
  \State \text{Initialize} {\raggedright #1}
}

\algnewcommand{\Create}[1]{%
  \State \text{Create} {\raggedright #1}
}

\algnewcommand{\Run}[1]{%
  \State \text{Run} {\raggedright #1}
}

\algnewcommand{\Compute}[1]{%
  \State \text{Compute} {\raggedright #1}
}

\algnewcommand{\Solve}[1]{%
  \State \text{Solve} {\raggedright #1}
}

\algnewcommand\algorithmicforeach{\textbf{for each}}
\algdef{S}[FOR]{ForEach}[1]{\algorithmicforeach\ #1\ \algorithmicdo}

\jyear{2023}%

\theoremstyle{thmstyleone}%
\theoremstyle{thmstyletwo}%

\theoremstyle{thmstylethree}%

\begin{document}

\title[Article Title]{Robust Monolithic versus Distributed Control/Structure Co-Optimization of Flexible Space Systems in Presence of Parametric Uncertainties}


\author*[1]{\fnm{Francesco} \sur{Sanfedino}}\email{francesco.sanfedino@isae.fr}

\author[1]{\fnm{Daniel} \sur{Alazard}}\email{daniel.alazard@isae.fr}

\author[2]{\fnm{Andy} \sur{Kiley}}\email{andy.kiley@airbus.com}
\author[2]{\fnm{Mark} \sur{Watt}}\email{mark.watt@airbus.com}

\author[3]{\fnm{Pedro} \sur{Simplicio}}\email{pedro.simplicio@ext.esa.int}
\author[3]{\fnm{Finn} \sur{Ankersen}}\email{finn.ankersen@ieee.org}

\affil*[1]{\orgname{ISAE-SUPAERO, Université de Toulouse}, \orgaddress{\street{10 Av. E. Belin}, \city{Toulouse}, \postcode{31055}, \country{France}}}

\affil[2]{\orgname{Airbus DS UK}, \orgaddress{\street{Gunnels Wood Rd}, \city{Stevenage}, \postcode{SG1 2AS}, \country{United Kingdom}}}

\affil[3]{\orgname{ESA-ESTEC}, \orgaddress{\street{Keplerlaan 1}, \city{Noordwijk}, \postcode{2201 AZ}, \country{The Netherlands}}}


\abstract{This paper presents an end-to-end framework for robust structure/control optimization of an industrial benchmark. When dealing with space structures, a reduction of the spacecraft mass is paramount to minimize the mission cost and maximize the propellant availability. However, a lighter design comes with a bigger structural flexibility and the resulting impact on control performance. Two optimization architectures (distributed and monolithic) are proposed in order to face this issue. In particular the Linear Fractional Transformation (LFT) framework is exploited to formally set the two optimization problems by including parametric uncertainties. Large sets of uncertainties have to be indeed taken into account in spacecraft control design due to the impossibility to completely validate structural models in micro-gravity conditions with on-ground experiments and to the evolution of spacecraft dynamics during the mission (structure degradation and fuel consumption). In particular the Two-Input Two-Output Port (TITOP) multi-body approach is used to build the flexible dynamics in a minimal LFT form.
The two proposed optimization algorithms are detailed and their performance are compared on an ESA future exploration mission, the ENVISION benchmark. With both approaches, an important reduction of the mass is obtained by coping with the mission's control performance/stability requirements and a large set of uncertainties.}


\maketitle

\section{Introduction}\label{sec1}

The widespread approach for Multi-Disciplinary Optimization (MDO) problems adopted in the space industry generally follows a sequential logic by neglecting the interconnection among different disciplines. However, since the optimization objectives in the different fields are often conflicting, this methodology can fail to find global optimal solutions. By restricting the analysis to just structure and control fields, the common hierarchy is to preliminarily define the structure by optimizing the physical design parameters and then leave the floor to the control optimization. This process can be iterated several times before a converging solution is found and control performance is met. Especially for large flexible structures, the minimization of the structural mass corresponds in fact to an increase in spacecraft flexibility, by bringing natural modes to lower frequencies where the interaction with the Attitude and Orbit Control System (AOCS) can be critical, especially in the presence of system uncertainties (\cite{FALCOZ201313}). Modern MDO techniques nowadays represents a tool to enhance the optimization task by integrating in a unique process all the objectives and constraints coming from each field.
Two kinds of architectures can be distinguished in the MDO framework: \textit{monolithic} and \textit{distributed} (\cite{martins2013multidisciplinary}). In a monolithic approach, a single optimization problem is solved, while in a distributed architecture the same problem is partitioned into multiple subproblems containing smaller subsets of the variables and constraints. Monolithic architectures have been proven to be more efficient  (\cite{Fathy2001}, \cite{Reyer2001}), than classical sequential strategies (\cite{chen20063d}, \cite{Li2001}), especially when bidirectional coupling exists between the two sub-problems (control and structure), for instance, when each of the two objectives depends on some common variables and parameters of each sub-problem (\cite{frischknecht2011pareto}).

In literature, several examples are found where control/structure co-design is implemented in a monolithic architecture. \cite{zhao2009control} presented a control–structural design optimization for vibration of piezoelectric intelligent truss structures. \cite{Allison2014} extended the direct transcription method, which transforms infinite-dimensional control design problems into finite-dimensional nonlinear programming problems, for co-design using a new automotive active suspension design example. \cite{MARANIELLO20161} presented an optimal vibration control and co-design strategy for very flexible actuated structures. A standard quasi-Newton method, the Sequential Least SQuares Programming (SLSQP) optimization algorithm   has been used to solve both the nonlinear optimal control problem and the co-design optimization. The implementation is monolithic and uses finite differences for the gradient evaluation. \cite{feng2014control} presented a multi-objective design for flexible spacecraft using a multiobjective evolutionary algorithm based on decomposition (MOEA/D) to reduce the total mass and optimize the control performance of a flexible spacecraft.
In \cite{alavi2021simultaneous} a Variable Neighborhood Search (VNS) metaheuristic method is developed to both minimize the structural mass and controlled system energy of a seismic civil structure. However, in all these works the uncertainties of the system are not taken into account.

The development in last decade of structured $\mathcal H_\infty$ control synthesis (\cite{GAHINET20111435}) opened the possibility to robust optimal co-design of structured controllers and tunable physical parameters. Linear Fractional Transformation (LFT) formalism allows in fact to embed in the dynamic model tunable physical parameters treated as parametric uncertainties. In addition, thanks to these techniques, particular properties can be imposed to the controller, as internal stability or performance respecting a frequency template, in the face of all the parametric uncertainties of the plant. This point is particularly important for aerospace applications where requirements are generally highly demanding and structural uncertainty, coming for example from an imperfect manufacturing or assembling, cannot be neglected. \cite{alazard2013avionics} demonstrated how this multi-model methodology implemented in $\mathcal{H}_\infty$ framework can be enlarged to include integrated design between certain tunable parameters of the controlled system and the stabilized structured controller. 
This approach has been used by \cite{PEREZ2015275} to optimize the structure of a deployable boom from TARANIS microsatellite, while meeting some control requirements. 

There exists as well in literature a large class of problems where coupling between structure and control is considered unidirectional (\cite{frischknecht2011pareto}). This means that the objective function $J_s (\mathbf y_s)$ of the structural sub-problem depends only on the structural design parameters $\mathbf y_s$ while the control criterion $J_c (\mathbf y_s,\mathbf y_c)$ depends on both structural ($\mathbf y_s$) and control ($\mathbf y_c$) design parameters, so that the system design objective becomes (\cite{frischknecht2011pareto}):
\begin{equation}
    J = w_sJ_s(\mathbf{y_s})+w_cJ_c(\mathbf{y}_s,\mathbf{y}_c) 
    \label{eq:sim_opt}
\end{equation}

Where $w_s$ and $w_c$ are ponderation weights. A partition of the structure and controller design variables is desirable for practical implementation when the impact of the controller variables on the structural objective is relatively small or computational means are not available to treat simultaneously control and structure variables in the objective function (\cite{frischknecht2011pareto}). A strategy in the latter case suggested by \cite{Fathy2001} and \cite{Reyer2001} is to solve the system-level problem as a nested optimization one, where the system solution is found with respect to $\mathbf y_s$, with the optimal $\mathbf y_c$ computed as function of $\mathbf y_s$ by solving the inner optimal controller problem first. This nested problem formulation is distinguished from the simultaneous one in Eq. \eqref{eq:sim_opt}:
\begin{equation}
    J^n = w_sJ_s(\mathbf{y_s})+w_cJ_c^n(\mathbf{y}_c^*(\mathbf{y}_s)) 
    \label{eq:nested}
\end{equation}
For this kind of problem, a distributed optimization architecture is then more appropriate.
There are several works in the literature, where nested optimization is used. \cite{Chilan2017} proposed a strain-actuated solar array concept that enables attitude slewing maneuvers and precision pointing stares for image acquisition, whilst simultaneously suppressing structural vibrations. \cite{zheng2021integrated} proposed a two layers integrated design optimization of actuator layout and structural ply parameters for the dynamic shape control of piezoelectric laminated curved shell structures. 

Two other examples in literature of distributed control/structure optimization were provided by the BIOMASS test case (\cite{FALCOZ201313}, \cite{toglia2013optimal}). In both works, a Genetic Algorithm was used to solve the global optimization problem and robust control techniques were applied for the nested control optimization problem. In the first work, limitations were encountered in the robust synthesis by considering the set of parametric uncertainties and an approximated dynamical uncertainty was instead used for the control synthesis and analysis. A conservative design is generally issued from this procedure and a small reduction of the spacecraft mass was finally obtained. Similar performance is obtained in \cite{toglia2013optimal}, where the controller synthesis does not take into account uncertainties and a formal $\mu$-analysis is run at each nested iteration in order to validate its robustness. The inconvenience in this approach is twofold: the computational time to run a $\mu$-analysis is high, especially in the presence of highly repeated parametric uncertainties (no information on the time performance is provided in the article); not considering uncertainties directly in the controller synthesis can invalidate a large number of controllers.

For the present study, both monolithic and distributed architectures are investigated on a real benchmark, the ENVISION spacecraft preliminary design.
In particular, the problem formulation in the multi-body Two-Input Two-Output Ports (TITOP) (\cite{alazard2015two}) modeling approach introduced by \cite{alazard2015two} allows the authors to easily define an MDO problem by including all possible system uncertainties from the very beginning of the spacecraft design. In this way, not only is a structure/control co-design possible, but system performance is robustly guaranteed.
The aim of this paper is to contribute to the evolution of industrial practice in robust control/structure co-design, by proposing a unified and generic approach based on a well-posed modeling problem that integrates both design parameters and parametric uncertainties in a unique representation. The advantage offered by this framework is twofold: to shortcut the unnecessary iterations among different fields of expertise and to speed up the validation and verification process by directly producing a robust preliminary design.

After recalling the principles of the multi-body TITOP modeling approach in Section \ref{sec:titop_theory} and showing how to build the uncertain plant of the ENVISION study case, the formulation of the robust control problem is presented in Section \ref{sec:robust_control}. Section \ref{sec:robust_control} details the proposed distributed and monolithic optimization algorithms and Section \ref{sec:results} presents the achieved results by comparing the two approaches and providing formal validation of the synthesized controllers. Section \ref{sec:conclusions} finally summarized all paper's contributions.

\section{Parametric Multi-Body Modeling}

When dealing with robust optimization of control and structural parameters of complex systems, a rigorous modeling framework able to take into account both optimization variables and system uncertainties is paramount. In the following sections the TITOP approach is presented with its direct application to the study case.

\subsection{Two-Input Two-Output Port Theory}
\label{sec:titop_theory}

\begin{figure}[th!]
	\centering
	\includegraphics[width=\columnwidth]{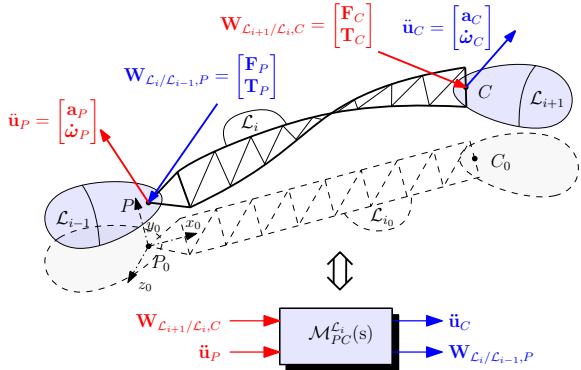}
	\caption{i-th flexible appendage sub-structured body (top) and equivalent TITOP model (bottom)}
	\label{fig:titop_beam}
\end{figure}

Let's consider the generic flexible appendage $\mathcal{L}_i$ in Fig. \ref{fig:titop_beam} linked to a parent substructure $\mathcal{L}_{i-1}$ at point $P$ and to a child substructure $\mathcal{L}_{i+1}$ at point $C$. 
Moreover let us define the reference frame $\mathcal{R}_0 = (P, x_0, y_0, z_0)$ centered in node P of $\mathcal{L}_i$ in equilibrium condition.
In the model of the appendage $\mathcal{L}_i$, clamped-free boundary conditions are considered: the joint at point P is rigid and statically determinate, with the parent body
$\mathcal{L}_{i-1}$ imposing a motion on $\mathcal{L}_i$, while point $C$ is internal and unconstrained, and the action of $\mathcal{L}_{i+1}$ is by means of a transmitted effort.

 The TITOP model $\mathcal{M}_{PC}^{\mathcal{L}_i}(\mathrm{s})$ is a linear state-space model with 12
inputs (6 for each of the two input ports):

\begin{enumerate}
	\item The 6 components in $\mathcal{R}_0$ of the wrench $\mathbf{W}_{\mathcal{L}_{i+1}/\mathcal{L}_i,C}$ composed of the three-components
	force vector $\mathbf{F}_C$ and the three-components torque vector $\mathbf{T}_C$ applied by $\mathcal{L}_{i+1}$ to $\mathcal{L}_i$ at
	the free node $C$;
	\item The 6 components in $\mathcal{R}_0$ of the acceleration vector $\ddot{\mathbf{u}}_P$ composed of the three-components linear acceleration vector $\mathbf{a}_P$ and the three-components angular acceleration vector $\dot{\bm{\omega}}_P$
	at the clamped node $P$;
\end{enumerate}

\noindent and 12 outputs (6 for each of the two output ports):

\begin{enumerate}
	\item The 6 components in $\mathcal{R}_0$ of the acceleration vector $\ddot{\mathbf{u}}_C$ at the free node $C$;
	\item The 6 components in $\mathcal{R}_0$ of the wrench $\mathbf{W}_{\mathcal{L}_i/\mathcal{L}_{i-1},P}$ applied by $\mathcal{L}_i$ to the parent structure
	$\mathcal{L}_{i-1}$ at the clamped node $P$.
\end{enumerate}

The TITOP model $\mathcal M^{\mathcal L_i}_{P,C}(\mathrm s)$ displayed in Fig. \ref{fig:titop_beam} (right) includes in a minimal state-space model the direct dynamic model (transfer from acceleration twist to wrench) at point $P$ and the inverse dynamic model (transfer from wrench to acceleration twist) at point $C$.

This model, conceived with the clamped-free condition, is useful to study any other kind of boundary configuration as proven by \cite{lin_dyn_flex} thanks to the invertibility of all of its 12 input-output channels.

The TITOP model $\mathcal M^{\mathcal L_i}_{P,C}(\mathrm s)$ can be then obtained analytically for simple geometries like beams (\cite{lin_dyn_flex}) and mechanisms (\cite{sanfedino2022integrated}) or numerically by Finite Element Model (FEM) analysis (\cite{sanfedino2018finite}). In the latter case, by considering the generalized modal coordinates $\bm{\eta}$ of the classical Craig-Bampton approach,  the state-space representation of $\mathcal M^{\mathcal L_i}_{P,C}(\mathrm s)$ is given by:
\begin{equation}
	\left[\begin{array}{c} 
		\dot{\bm{\eta}} \\ \ddot{\bm{\eta}} \\ \hline
		\ddot{\mathbf{u}}_C \\ \mathbf{W}_{\mathcal{L}_{i}/\mathcal{L}_{i-1},P}
	\end{array} \right] =
	\left[\begin{array}{c|c}
		 \mathbf{A} & \mathbf{B} \\
		 \hline
		 \mathbf{C} & \mathbf{D}
	\end{array} \right]
	\left[\begin{array}{c} 
		\bm{\eta} \\ \dot{\bm{\eta}} \\ \hline \mathbf{W}_{\mathcal{L}_{i+1}/\mathcal{L}_{i},C} \\ \ddot{\mathbf{u}}_P
	\end{array} \right],
 \label{eq:titop_fem}
\end{equation}
where
\begin{equation*}
	\begin{array}{c}
		\mathbf{A} = \left[\begin{array}{cc}
\mathbf{0}_{N_i\times N_i} & \mathbf{I}_{N_i} \\
-\mathbf{k} & -\mathbf{c}
\end{array}\right],\quad\quad \mathbf{B} = \left[\begin{array}{cc}
\mathbf{0}_{N_i\times 6} & \mathbf{0}_{N_i\times 6} \\
\mathbf{\Phi}_{C}^{\mathrm{T}} & -\mathbf{L}_{P}
\end{array}\right], \\
\mathbf{C}= \left[\begin{array}{cc}
	-\mathbf{\Phi}_{C}\mathbf{k} & -\mathbf{\Phi}_{C}\mathbf{c} \\
	\mathbf{L}_{P}^{\mathrm{T}}\mathbf{k} & \mathbf{L}_{P}^{\mathrm{T}}\mathbf{c}
\end{array}\right],  \\
\mathbf{D} = \left[\begin{array}{cc}
	\mathbf{\Phi}_{C}\mathbf{\Phi}_{C}^{\mathrm{T}} & (\bm{\tau}_{CP}-\mathbf{\Phi}_{C}\mathbf{L}_{P}) \\
	(\bm{\tau}_{CP}-\mathbf{\Phi}_{C}\mathbf{L}_{P})^{\mathrm{T}} & \mathbf{L}_{P}^{\mathrm{T}}\mathbf{L}_{P} - \mathbf{M}_{\mathrm{rr}}
\end{array}\right].
\end{array}
\end{equation*}

\begin{figure*}[!t]
    \centering
    \includegraphics[width=\textwidth]{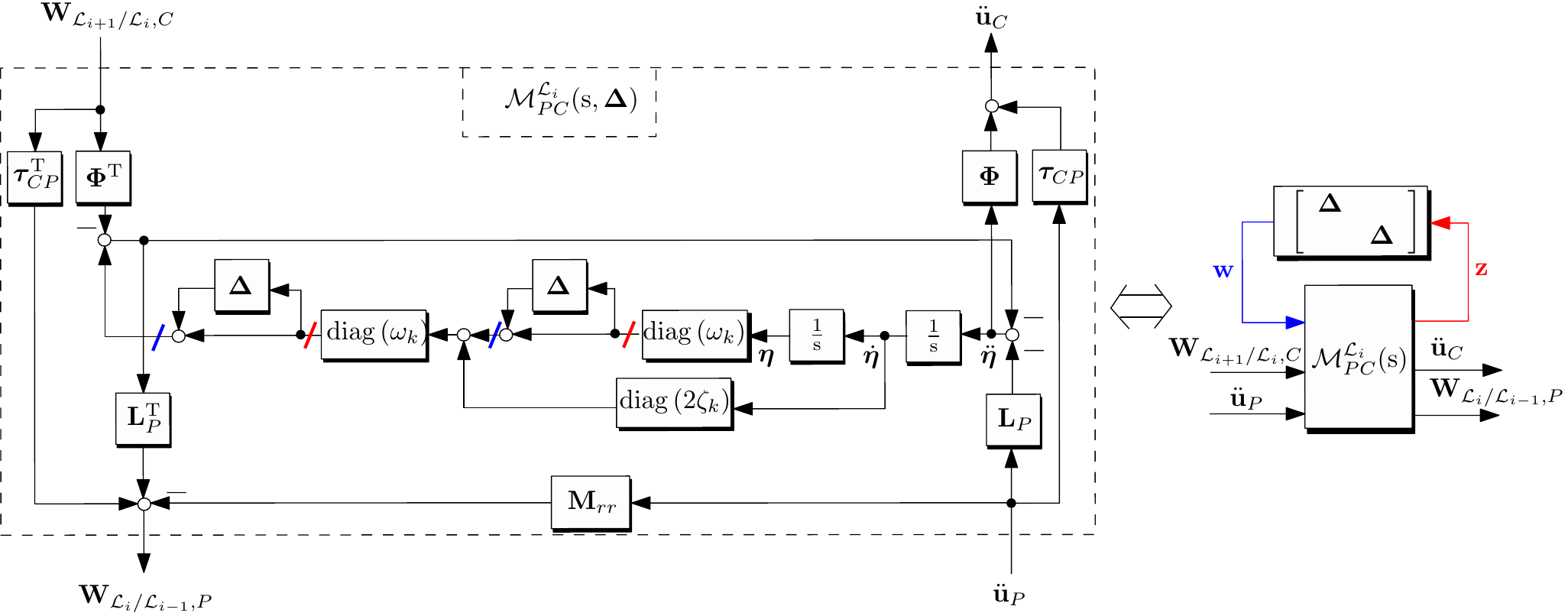}
    \caption{Uncertain TITOP model of a generic flexible structure based on FEM modal analysis (left) and its equivalent LFT form (right)}
    \label{fig:NASTRAN_block_diagram}
\end{figure*}

\begin{itemize}
	\item $\mathbf{k}=\mathrm{diag}(\omega_k^2)$: is the diagonal matrix of the square value of the retained $N_\omega$ frequencies of the flexible modes $\omega_k$, with $k\in 1\dots N_\omega$;
	\item $\mathbf{c}=\mathrm{diag}(2\zeta_k\omega_k)$: damping matrix, where $\zeta_k$ is the modal damping factor associated to mode $\omega_k$;
	\item $\mathbf{L}_P$: matrix of the participation factors w.r.t. $P$;
	\item $\mathbf{\Phi}_C$: projections of the flexible mode shapes on $C$;
	\item $\mathbf{M}_{rr} =\mathbf{M}_{r} - \mathbf{L}_P^\mathrm{T}\mathbf{L}_P $: residual mass, obtained from the rigid body mass matrix $\mathbf{M}_{r}$;
        \item $\bm{\tau}_{CP}$ describes the rigid kinematic model between the DOFs of the generic internal node $C$ and the junction DOFs of the node $P$. For a clamped (in $P$) - free (in $C$) flexible structure:
        \begin{equation}
	           \bm{\tau}_{CP} = \begin{bmatrix}\mathbf{I}_{3} & ^{*}\overrightarrow{\mathbf{{CP}}} \\ \mathbf{0}_{3} & \mathbf{I}_{3} \end{bmatrix},
        \end{equation}
        where $^{*}\overrightarrow{\mathbf{{CP}}}$ is the skew-symmetric matrix associated with the vector from $C$ to $P$ of the flexible appendage in the undeformed configuration.
\end{itemize}
All the parameters needed to build the model \eqref{eq:titop_fem} can be recovered by running a modal analysis with a commercial software like MSC NASTRAN. 
A particularity of TITOP models is to have access to each structural parameter in an analytical way and have the possibility to easily assign it an uncertainty. In this way the model can be straightforward put in a generalized LFT form, directly exploitable for modern robust control synthesis/analysis techniques.
If in fact we consider for instance that now in model \eqref{eq:titop_fem} at each modal frequency $\omega_k$ is associated a parametric uncertainty $\delta\omega_k\in\left[-1,1\right]$ such that the uncertain frequency is now expressed as $\tilde{\omega}_k=\omega_k\left(1+\delta\omega_k\right)$, the corresponding TITOP model $\mathcal M^{\mathcal L_i}_{P,C}(\mathrm s,\bm{\Delta})$ (with $\bm{\Delta}=\mathrm{diag}(\delta\omega_k)$) can be represented by the block diagram as in Fig. \ref{fig:NASTRAN_block_diagram} (left side) with a minimal number of repetitions of the uncertainties. 
Note that parameters to be optimized can be isolated in the same way. 

In the same figure the equivalent LFT model is shown on the right side, where the uncertain blocks can be isolated with respect to the nominal plants thanks to the exogenous inputs $\mathbf{w}$ and outputs $\mathbf{z}$, obtained by opening $\mathcal M^{\mathcal L_i}_{P,C}(\mathrm s,\bm{\Delta})$ respectively in the points marked in red and blue in Fig. \ref{fig:NASTRAN_block_diagram} (left).

The TITOP approach was finally extended to multiple ports in case several child structures are connected to the $i$-th flexible sub-structure (\cite{sanfedino2018finite}).  

\subsection{Spacecraft Assembly}

\begin{figure*}[!t]
	\centering
	\includegraphics[width=\textwidth]{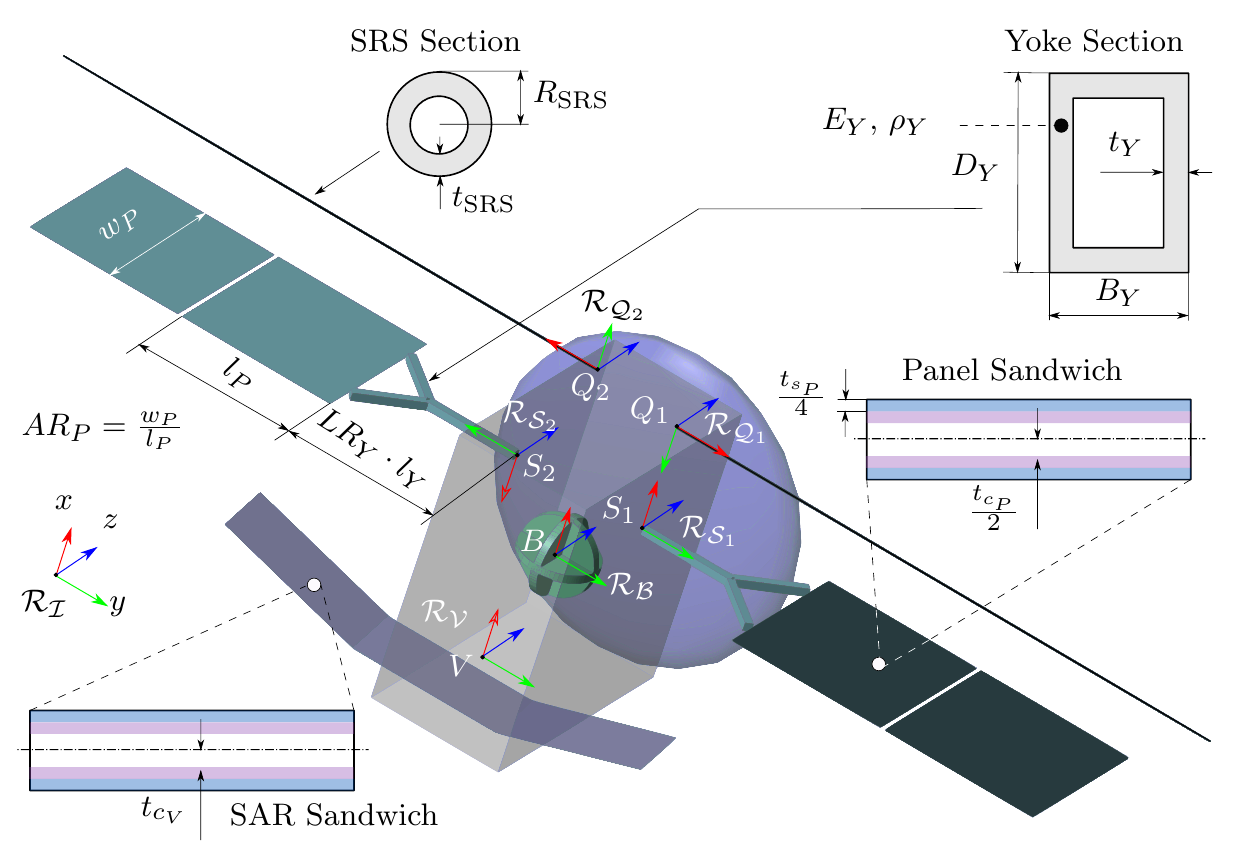}
	\caption{ENVISION spacecraft: definition of geometry and reference frames}
	\label{fig:EnvisionSpacecraft}
\end{figure*}

Once a TITOP model is obtained for each sub-structure of a multi-body system, the assembly is easily done by connecting the ports corresponding to the connection points among the sub-elements. Let's consider the ENVISION spacecraft depicted in Fig. \ref{fig:EnvisionSpacecraft}. It is constituted by a rigid central body $\mathcal{B}$, two solar arrays $\mathcal{S}_1$ and $\mathcal{S}_2$, a Subsurface Radar System (SRS) composed of two flexible beams $\mathcal{Q}_1$ and $\mathcal{Q}_2$, and a Synthetic Aperture Radar (SAR) $\mathcal{V}$. In Fig. \ref{fig:EnvisionSpacecraft} it can be noticed that the central body reference frame is centered in its center of mass (CoM) $B$, while all appendages have their body frame defined in correspondence of their attachment nodes. 

\begin{figure}[ht!]
    \centering
    \includegraphics[width=\columnwidth]{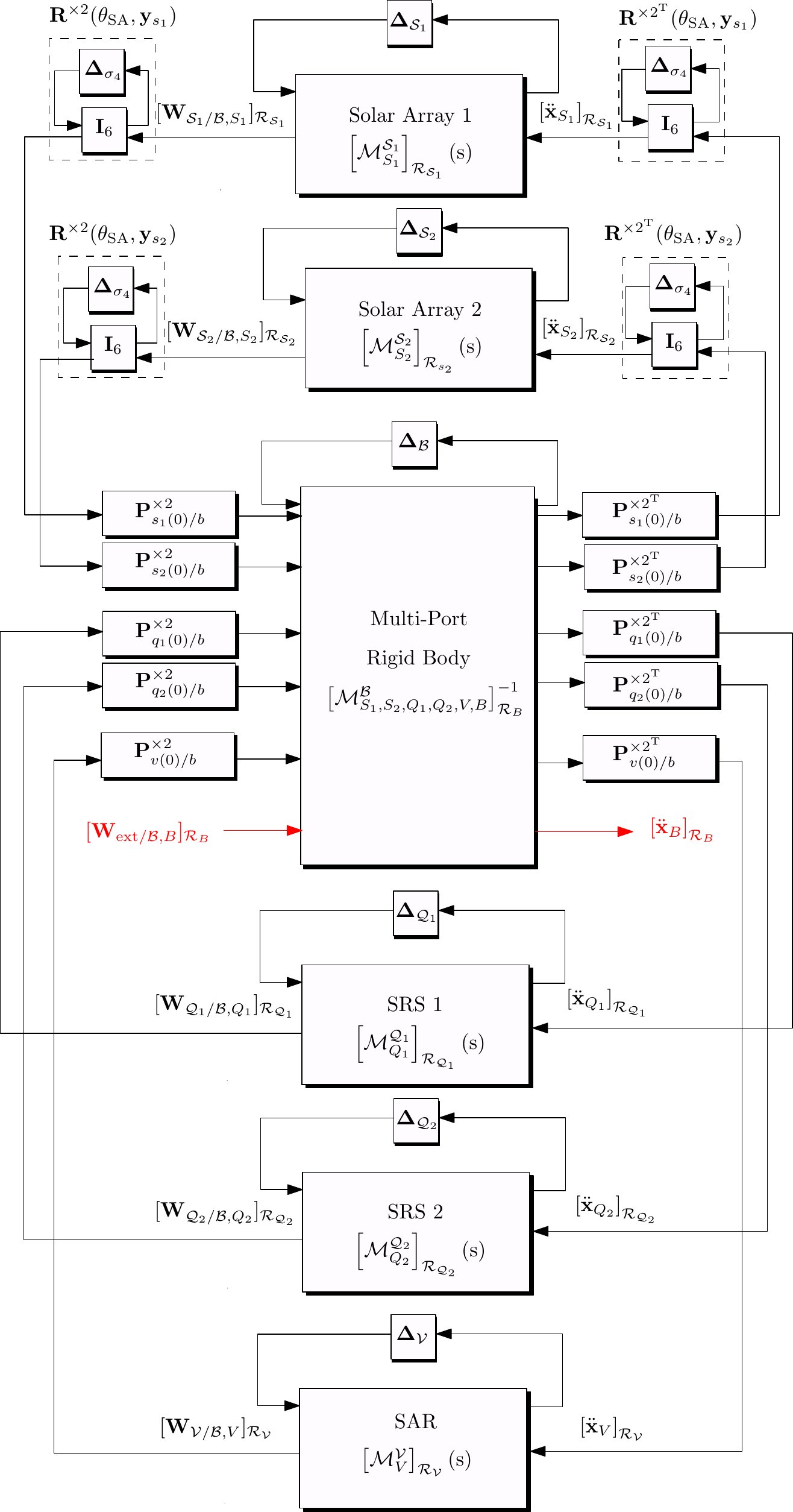}
    \caption{Assembled dynamics of the ENVISION spacecraft in TITOP framework}
    \label{fig:assembled_system}
\end{figure}

The assembled model of the whole spacecraft in TITOP framework is shown in Fig. \ref{fig:assembled_system}. Note that each sub-components has its corresponding LFT block connected to the other block throw wrench of forces/torques and linear/angular accelerations. In particular:
\begin{itemize}
    \item the central body $\mathcal{B}$ is modeled as a Multi-Port Rigid Body in which all inputs are the wrenches applied to all connection points by the appendages and the resultant wrench of the external perturbation forces/torques $\left[\mathbf{W}_{\mathrm{ext},\mathcal{B},B}\right]_{\mathcal{R}_B}$ applied to its CoM $B$. See \cite{sanfedino2019experimental} for more details on this model;
    \item the two solar panels, the two beams constituting the SRS antenna and the SAR antenna are modeled as generic flexible structures built from NASTRAN FEM analysis as discussed in Section \ref{sec:titop_theory}.
\end{itemize}

Note that in Fig. \ref{fig:assembled_system} the dynamical models of the appendages are connected to the main body structure through the rotation matrices $\mathbf{P}^{\times 2}_{a(0)/b}$ that express the wrench vectors expressed in the body frame of appendage $\mathcal{A}$ into the body frame of the main body $\mathcal{B}$. The transpose of these matrices allows projecting the acceleration vectors instead. Since the solar arrays are not fixed on the ENVISION platform, the rotation matrices $\mathbf{R}^{\times 2}(\theta_\mathrm{SA},\mathbf{y}_{s_\bullet})$, dependent on the solar array configuration angle $\theta_\mathrm{SA}$ around the rotation axis $\mathbf{y}_{s_\bullet}$, take int account the presence of a Solar Array Drive Mechanism (SADM) and can be easily expressed in a LFT way as shown in \cite{dubanchet2016modeling}, where the considered uncertain parameter is $\sigma_4 = \tan(\theta_\mathrm{SA}/4)$, such that $\bm\Delta_{\sigma_4} = \sigma_4\mathbf{I}_8$.
Finally all other uncertain blocks $\bm\Delta_\bullet$ take into account all possible uncertainties and optimization parameters associated to each substructure.

The main advantages of having the system expressed in TITOP framework are listed below:
\begin{itemize}
    \item physical understanding is conserved at sub-component level;
    \item parametric uncertainty can be considered at sub-component level;
    \item redesign of sub-component in preliminary design phase is easy;
    \item adaptability to a user-friendly Matlab/Simulink toolbox. The Satellite Dynamics Toolbox library (SDTlib) allows to easily model dynamical space systems in TITOP approach (\cite{sdt}).
\end{itemize}

\begin{figure*}[t!]
    \centering
    \includegraphics[width=\linewidth]{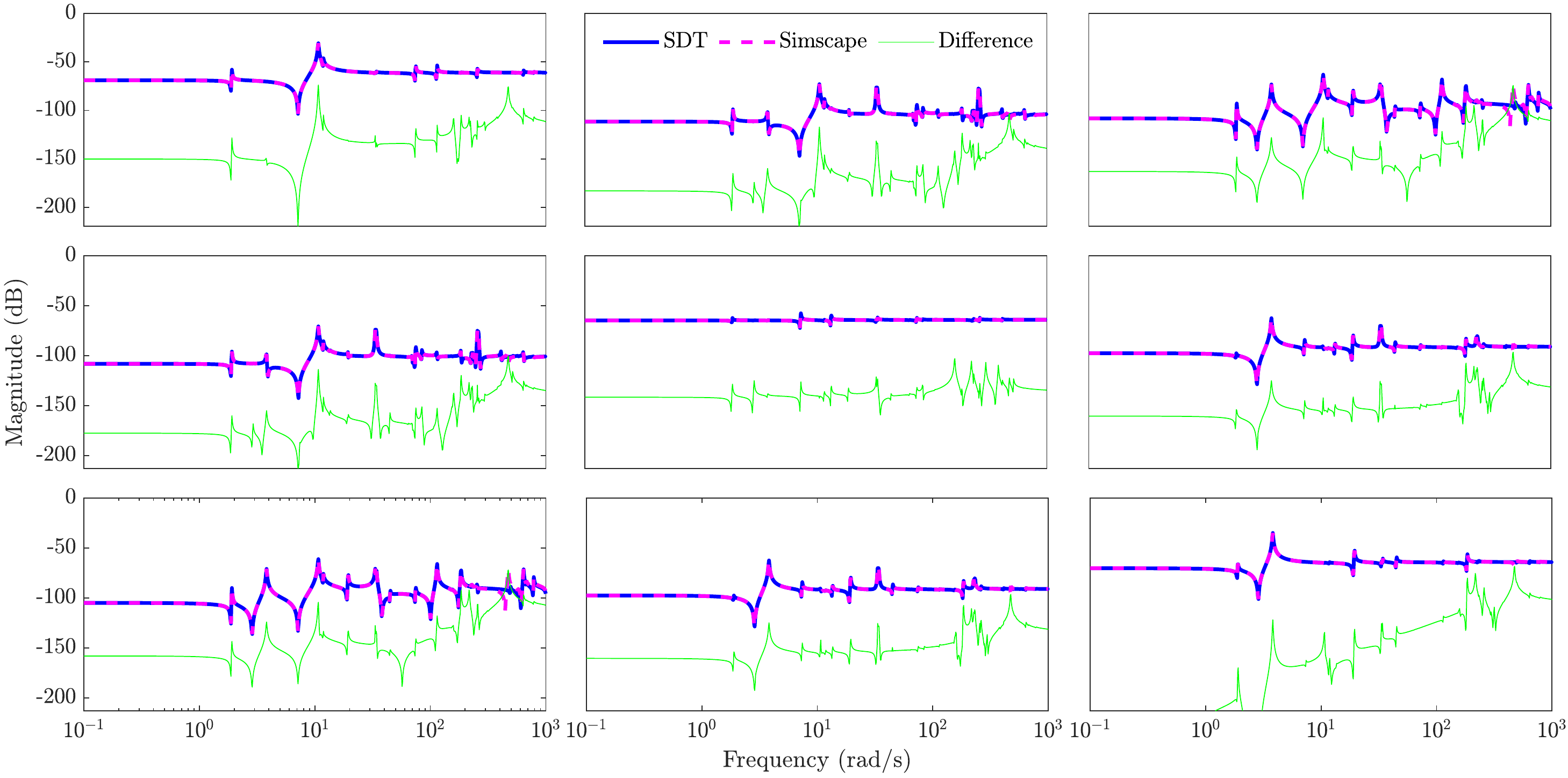}
    \caption{Comparison of SDT and Simscape models of the transfer function $\left[\mathbf{W}_{\mathrm{ext/\mathcal{B},B}}\right]_{\mathcal{R}_B}\left(4:6\right)\rightarrow \left[\ddot{\mathbf{x}}_B\right]_{\mathcal{R}_B}\left(4:6\right)$}
    \label{fig:comparisonSDTSimscape}
\end{figure*}

The assembled TITOP model is validated with an equivalent model built in Simscape, where Reduced Order Flexible Solid (ROFS) blocks are used for all flexible appendages. In order to correctly parameterize the constitutive mass, stiffness and damping matrices directly for a NASTRAN FEM analysis, the approach presented in \cite{alazard2023port} is used. A comparison of the $3\times 3$ transfer functions $\left[\mathbf{W}_{\mathrm{ext/\mathcal{B},B}}\right]_{\mathcal{R}_B}\left(4:6\right)\rightarrow \left[\ddot{\mathbf{x}}_B\right]_{\mathcal{R}_B}\left(4:6\right)$ from the three external torques applied at point $B$ to the three angular accelerations experienced by point $B$, is shown in Fig. \ref{fig:comparisonSDTSimscape}. Notice as SDTlib and Simscape models are close and their difference is several orders of magnitude smaller than their absolute values. This comparison is based on the same nominal configuration of the ENVISION plant, where all parametric variation are fixed to a particular value. 

The introduction of uncertainties in TITOP formalism then makes SDTlib model directly exploitable for robust control synthesis/analysis and monolithic optimization.  

\subsubsection{Open-loop parametric analysis}
\begin{figure}[!ht]
    \centering
    \includegraphics[width=\columnwidth]{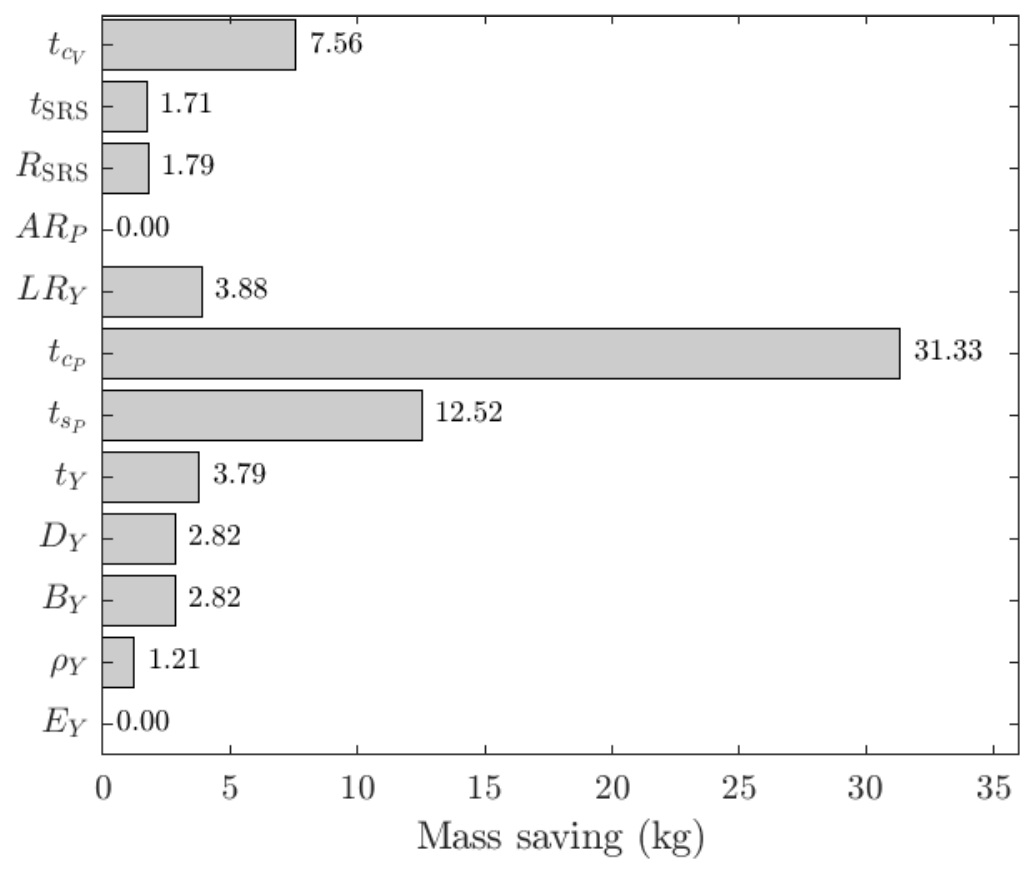}
    \caption{Potential mass saving (in kg) obtained by optimization of the ENVISION design parameters}
    \label{fig:MassSaving}
\end{figure}

\begin{figure*}[!ht]
\centering
     \includegraphics[width=\textwidth]{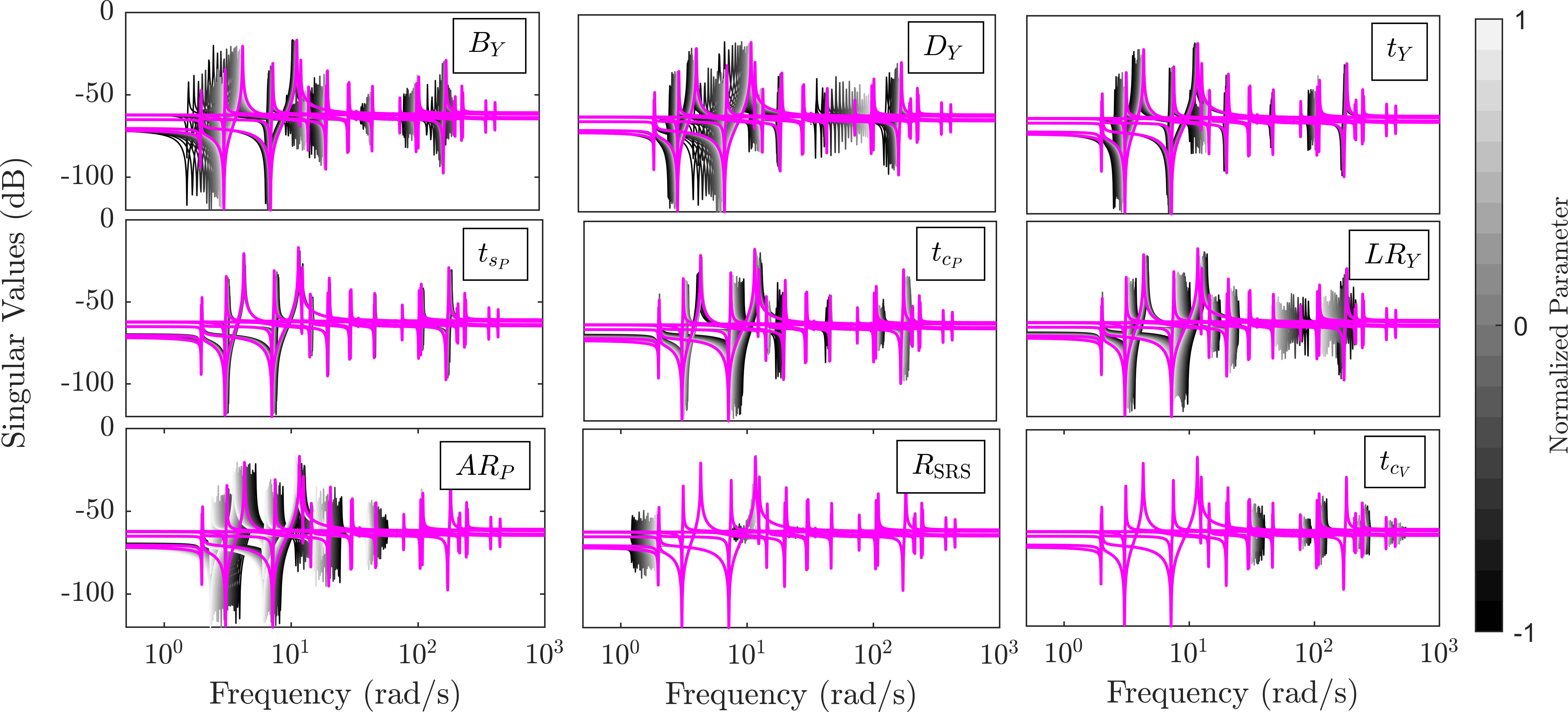}
        \caption{Evolution of the singular values of the transfer function $\left[\mathbf{W}_{\mathrm{ext/\mathcal{B},B}}\right]_{\mathcal{R}_B}\rightarrow \left[\ddot{\mathbf{x}}_B\right]_{\mathcal{R}_B}$ with variation of the optimization parameters}
        \label{fig:FreqSensitivity}
\end{figure*}

An advantage of having an LFT model is that a large family of possible plants is available in a continuous way in a unique representation. In this way, it is possible to see, for instance, which parameters will impact the mass the most. If the optimization parameters in Table \ref{tab:opt_param} are considered for the ENVISION benchmark, the potential gain in spacecraft overall mass after optimization is shown in Fig. \ref{fig:MassSaving}. 
In the context of co-design of structure and control architecture it is also important to see the impact of the design parameters on the system flexibility. Some natural modes can in fact interact with the control bandwidth by causing a significant degradation of the control stability/performance.
Figure \ref{fig:FreqSensitivity} shows the singular values of the transfer function $\left[\mathbf{W}_{\mathrm{ext/\mathcal{B},B}}\right]_{\mathcal{R}_B}\rightarrow \left[\ddot{\mathbf{x}}_B\right]_{\mathcal{R}_B}$ when one optimization parameter varies in its admissible range and all other optimization parameters impacting the mass are fixed to their maximum value. When not varying, the panel aspect ratio and the yoke length ratio are equal to unity and the Yoke Young Modulus takes its minimum value. Note that the situation in which all parameters are fixed is marked in magenta and no uncertainty is considered in this analysis.
Only the most impacting parameters provoking a shift of normal modes is depicted in Fig. \ref{fig:FreqSensitivity}: i.e. a reduction of the SRS outer radius $R_\mathrm{SRS}$ or the side lengths ($B_Y$ and $D_Y$) of the Yoke section make the first modes shift to lower frequencies. 

\begin{figure}[!ht]
    \centering
    \includegraphics[width=\linewidth]{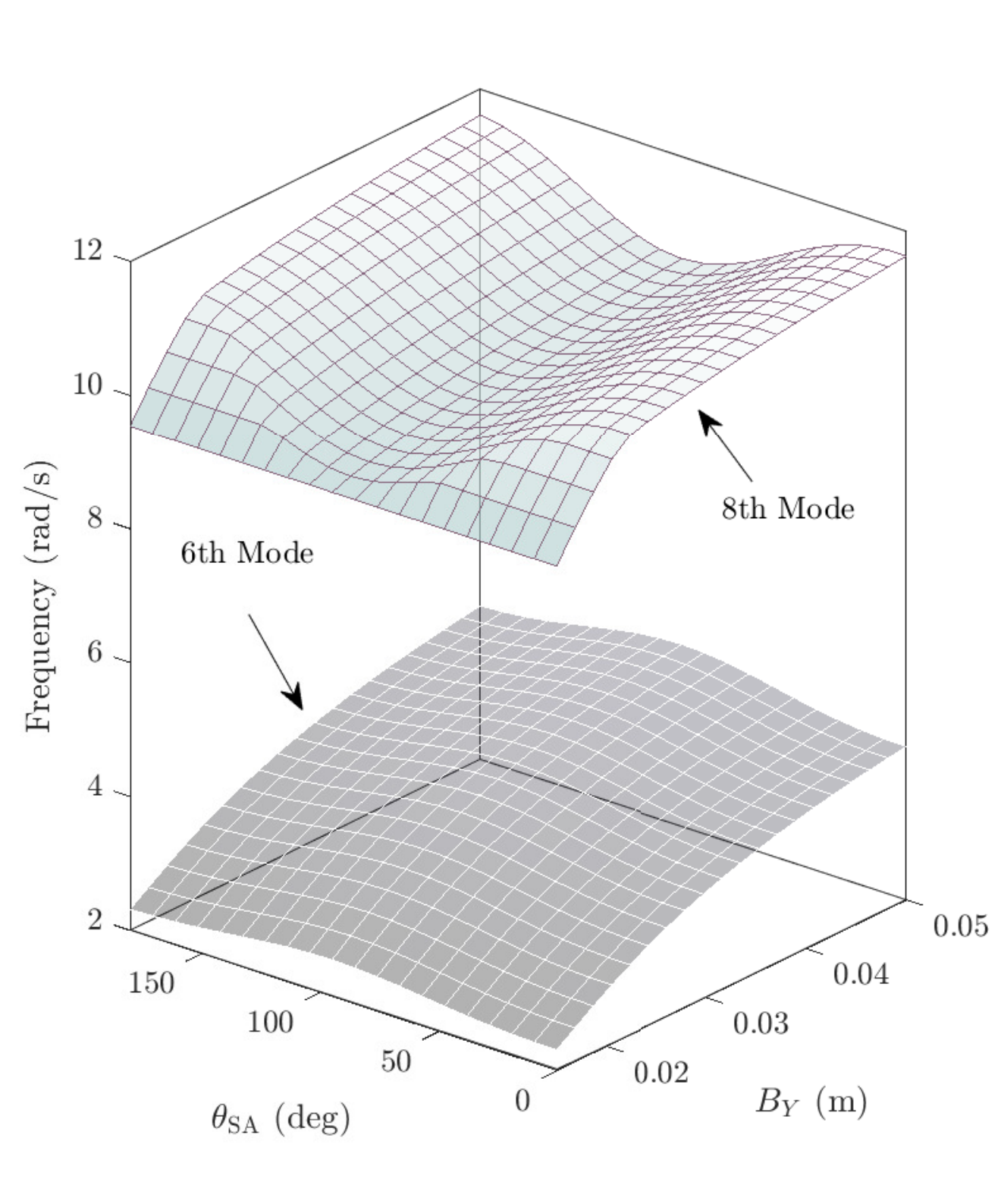}
    \caption{Evolution of 6th and 8th mode frequencies with variation of $B_Y$ and $\theta_\mathrm{SA}$}
    \label{fig:thetaByfreq}
\end{figure}

Another phenomenon captured by the TITOP modeling is the shift of natural modes caused by the rotation of the two solar arrays. Figure \ref{fig:thetaByfreq} shows for instance the evolution of the frequency of the 6th and 8th mode when both $B_Y$ and $\theta_\mathrm{SA}$ vary.   

\subsubsection{Modeling uncertainty for distributed and monolithic optimization}
\label{sec:delta_monomithic}

\begin{figure} [!h]
    \centering
    \includegraphics[width=\columnwidth]{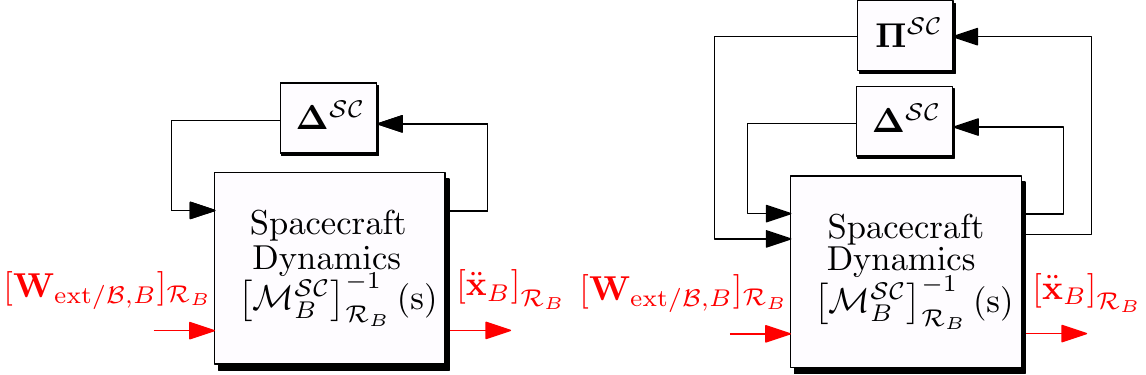}
    \caption{LFT model of the ENVISION spacecraft used for distributed (left) and monolithic (right) optimization}
    \label{fig:LFT_SC}
\end{figure}

When dealing with distributed optimization, the decoupling between the global structural problem from the nested control optimization, shown in Eq. \eqref{eq:nested}, requires that the optimization structural variables $\mathbf{y}_s$ are fixed during control synthesis. For this reason, in the distributed optimization the overall uncertain block $\bm\Delta^{\mathcal{SC}}$, obtained by putting the model depicted in Fig. \ref{fig:assembled_system} in the generalized LFT form, contains only the set of real uncertainties listed in Table \ref{tab:uncertainty_list}. When a monolithic optimization routine is chosen instead, an uncertain block $\bm\Pi^{\mathcal{SC}}$ containing the optimization variable is considered as well. In the monolithic approach in fact, as shown in Section \ref{sec:robust_control}, the non-smooth optimization routine used for robust control synthesis (\cite{apk2007}) is able to provide the optimized control and structural parameters at the same time by coping with all considered system uncertainties.
The two different LFT models are shown in Fig. \ref{fig:LFT_SC}.

\begin{table}[!h]
    \centering
    \resizebox{\columnwidth}{!}{\begin{tabular}{cccc}
    \hline
       \textbf{Parameter}  & \textbf{Description} & \textbf{Occurrence} & \textbf{Uncertainty}  \\ \hline
        $m^\mathcal{B}$ & Mass $\mathcal{B}$ & 3 & $\pm 15\%$ \\
        $I_{xx}^\mathcal{B}$ & Inertia x-axis $\mathcal{B}$ & 1 & $\pm 15\%$ \\
        $I_{yy}^\mathcal{B}$ & Inertia y-axis $\mathcal{B}$ & 1 & $\pm 15\%$ \\
        $I_{zz}^\mathcal{B}$ & Inertia z-axis $\mathcal{B}$ & 1 & $\pm 15\%$ \\
        $\omega^{\mathcal S_\bullet}_1$ & 1st mode $\mathcal{S}_\bullet$ & 4 & $\pm 25\%$ \\
        $\omega^{\mathcal S_\bullet}_2$ & 2nd mode $\mathcal{S}_\bullet$ & 4 & $\pm 25\%$ \\
         $\omega^{\mathcal V}_1$ & 1st mode $\mathcal{V}$ & 2 & $\pm 25\%$ \\
        $\omega^{\mathcal V}_2$ & 2nd mode $\mathcal{V}$ & 2 & $\pm 25\%$ \\
        $\sigma_4$ & Configuration $\mathcal{S}_\bullet$ & 32 & $\left[-1\quad1\right]$ \\ \hline
    \end{tabular}}
    \caption{Parametric uncertainties of the ENVISION spacecraft}
    \label{tab:uncertainty_list}
\end{table}

\begin{figure*}[!ht]
    \centering
    \includegraphics[width=\linewidth]{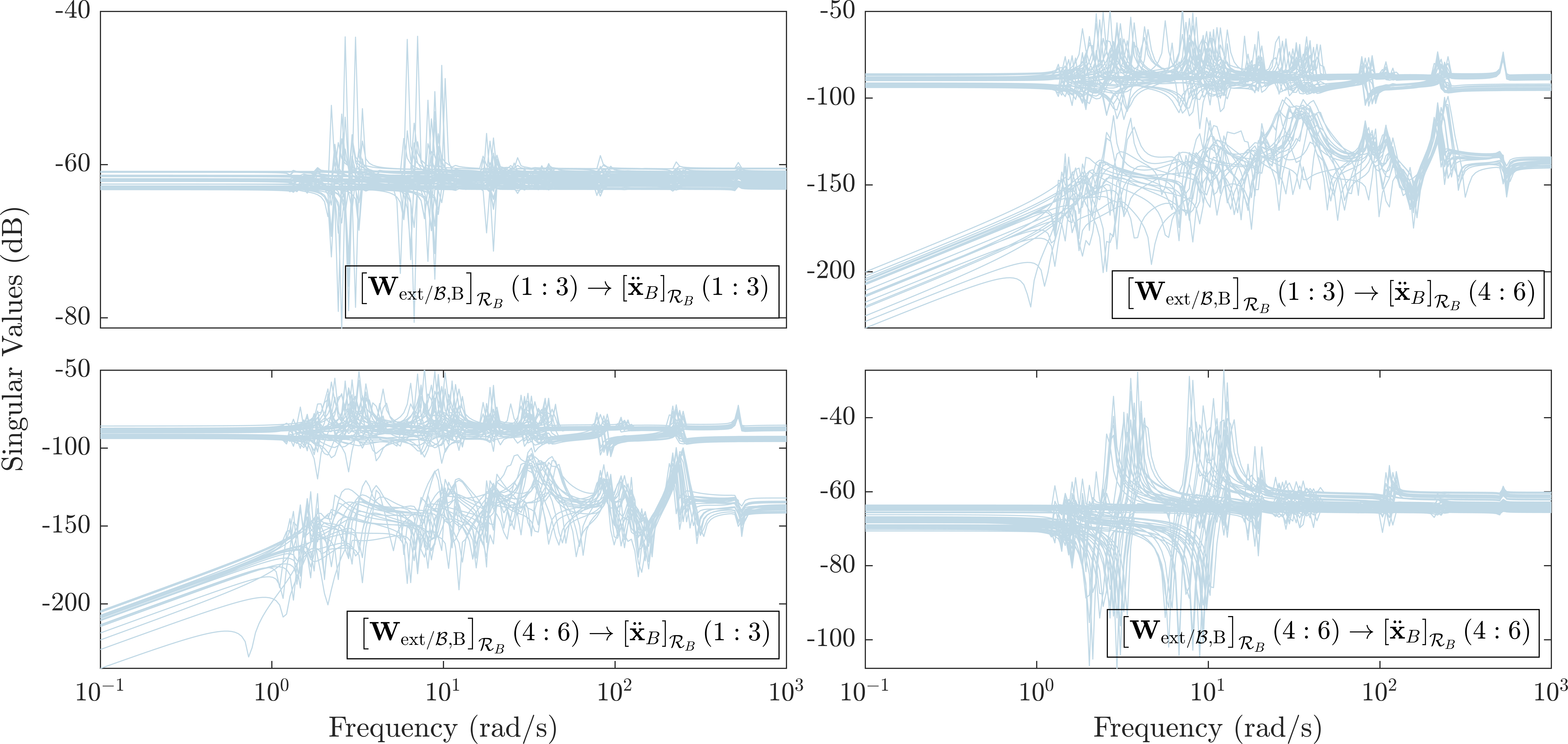}
    \caption{LFT model of the ENVISION plant by considering all uncertainties and optimization parameters for monolithic optimization}
    \label{fig:LFT_APRICOT}
\end{figure*}

An important task to be accomplished before running a monolithic optimization is to correctly model the uncertain block $\bm\Pi^\mathcal{SC}$. The major difficulty in this task is to keep the number of uncertainty repetitions as small as possible in order to not run into numerical problem when the robust control synthesis is tackled. When using a complex FEM model, this number can grow very fast with the number of the nodes in the mesh. For this reason, only four optimization parameters are selected in the list in Table \ref{tab:opt_param}, mostly impacting the overall mass and the shift of the first modes to lower frequencies: panel skin $t_{s_P}$ and core thickness $t_{c_P}$, SRS section outer radius $R_\mathrm{SRS}$ and the SAR core thickness $t_{c_V}$.
The way adopted in this work to face this problem is to get a multivariate polynomial approximation $\bm\Pi^\mathcal{SC}$ by using a set of samples and solving a classical Linear Least-Squares problem as proposed by \cite{poussot2012generation}. The algorithm available in the \textit{lsapprox} routine of the APRICOT library \cite{roos2014polynomial} is used to get an approximation of the delta-blocks for the matrices $\mathbf{M}_{rr}$, $\mathbf{L}_p$ and $\mathrm{diag}(\omega_k)$ in Fig. \ref{fig:NASTRAN_block_diagram}. A limited number of models is needed to generate a very precise approximation (order of magnitude of maximum relative error equal to $10^{-2}$): 100 different NASTRAN models are generated with random values of $t_{s_P}$ and $t_{c_P}$ in their ranges and by including the four corner scenarios, 10 NASTRAN models are obtained for the SRS beam antennas by linearly spanning $R_\mathrm{SRS}$ and 10 different NASTRAN models of the SAR antenna are finally generated by linearly spanning $t_{c_V}$.
Once assembled, the $\bm\Pi^\mathcal{SC}$ block contains the following number of occurrences: 220 for $t_{s_P}$, 248 for $t_{c_P}$, 64 for $R_\mathrm{SRS}$ and 116 for $t_{c_V}$. Figure \ref{fig:LFT_APRICOT} shows the singular values of the final LFT model of the ENVISION spacecraft used for monolithic optimization by taking into account all uncertainties and optimization parameters. 

\section{Robust Attitude Control}
\label{sec:robust_control}

The objective of this section is to present the general control architecture of the ENVISION Attitude Control System (ACS). 

We consider the robust design of a $3$-axis  structured attitude control law to meet:
\begin{itemize}
\item (\texttt{\textbf{Req1}}) the absolute pointing requirement, defined by the $3\times 1$ vector of Absolute Performance Error (\textbf{APE} $=\left[0.08\quad0.2\quad0.08\right]^\mathrm{T}\cdot10^{-3}$ rad), in spite of low frequency orbital disturbances dominated by the gravity gradient torque (characterized by the $3\times 1$ upper bound on the magnitude $\mathbf{T}_\mathrm{ext}=\left[1.9\quad1.9\quad 1.9\right]^\mathrm{T}\cdot10^{-3}$ Nm), 
\item (\texttt{\textbf{Req2}}) the relative pointing requirement, defined by the $3\times 1$ vector of Relative Performance Error (\textbf{RPE} $=\left[0.5\quad0.5\quad0.5\right]^\mathrm{T}\cdot10^{-3}$ rad) to be kept for a time window $\Delta t_\mathrm{RPE} =15\,\mathrm{s}$,
\item (\texttt{\textbf{Req3}}) the maximum command requirement, defined by the $3\times 1$ vector of maximum control torque ($\bar{\mathbf u}=\left[0.215\quad0.215\quad0.215\right]^\mathrm{T}$ Nm),
\item (\texttt{\textbf{Req4}}) stability margins characterized by an upper bound $\gamma$ on the $\mathcal H_{\infty }$-norm of the input sensitivity function,
\end{itemize}
while minimizing the variance of the APE and RPE in response to the star sensor and gyro noises characterized by their Power Spectral Density (PSD), respectively \textbf{PSD}$^\mathrm{SST}=(3.5\cdot10^{-5})^2\mathbf{I}_3\,\mathrm{rad^2 s}$ and \textbf{PSD}$^\mathrm{GYRO}=(1.4\cdot10^{-6})^2\mathbf{I}_3\,\mathrm{rad^2/s}$ (assumed to be equal for the $3$ components).

The value $\gamma =1.5$ ensures on each of the 3 axes:
\begin{itemize}
\item a disk margin $>1/\gamma =0.667$,
\item a gain margin $>\frac{\gamma }{\gamma -1}=3\,(9.542\,\mathrm{dB})$,
\item a phase margin $>2\arcsin \frac{1}{2\gamma }=38.9\,\mathrm{deg}$. 
\end{itemize}

The requirements \texttt{\textbf{Req1}}, \texttt{\textbf{Req2}} and \texttt{\textbf{Req3}} must be met for any values of the uncertain mechanical parameters regrouped in the block $\bm{\Delta}^\mathcal{SC}$.

\begin{figure*}[ht!]
    \centering
    \includegraphics[width=\textwidth]{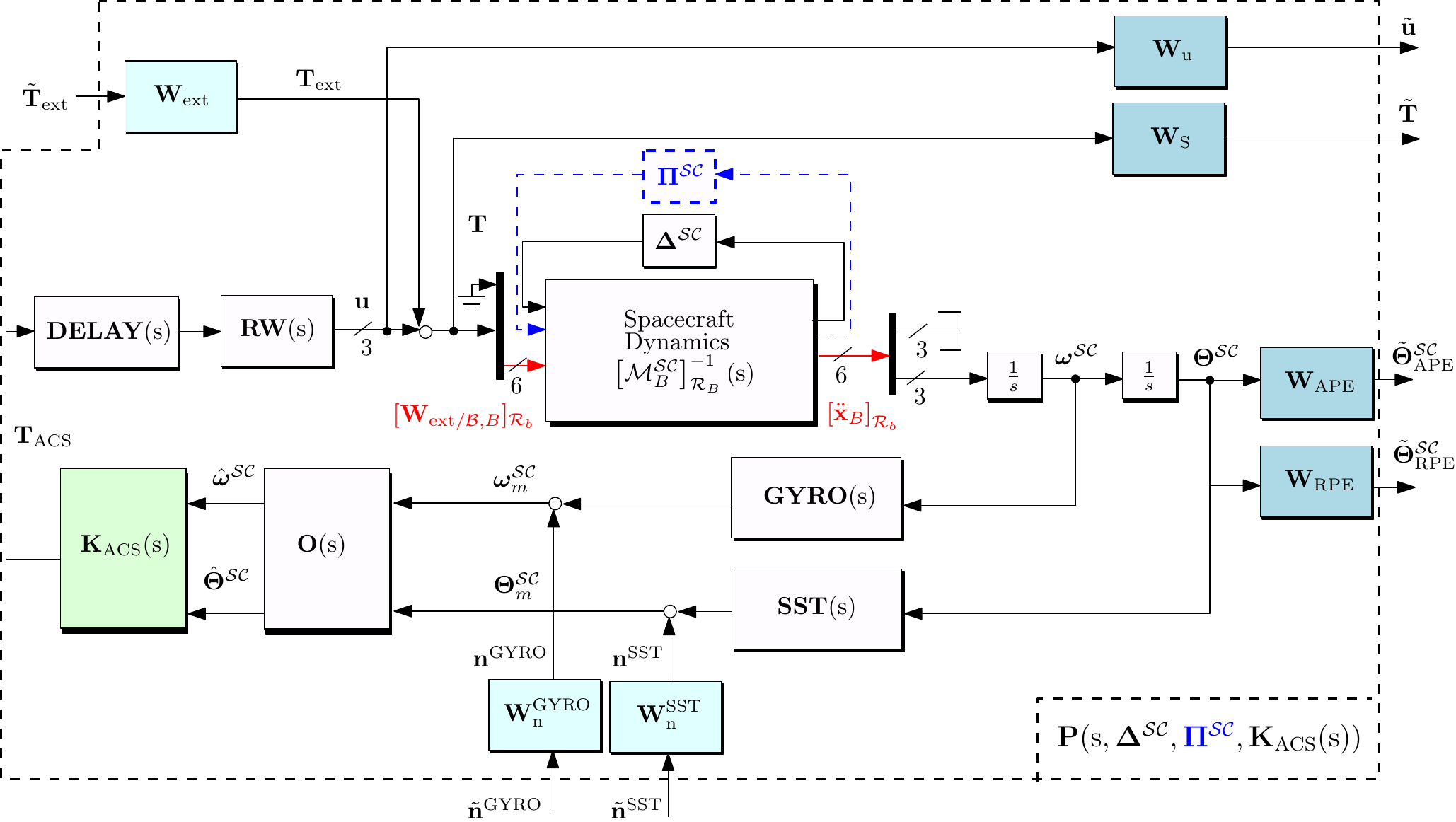}
    \caption{Attitude control system architecture}
    \label{fig:AOCS}
\end{figure*}

By referring to Fig. \ref{fig:AOCS}, the models of the avionics components are: 
\begin{itemize}
    \item A reaction wheel system modeled as a second order dynamics with cut-off natural frequency of $100\pi$ rad/s and damping factor 0.7:
        \begin{equation}
            \textbf{RW}(\mathrm s)=\frac{(100\pi)^2}{\mathrm s^2+140\pi\mathrm{s}+(100\pi)^2}\mathbf{I}_3
        \end{equation}
    \item A gyro sensor modeled as a first-order dynamics with cut-off frequency $200\pi$ rad/s:
        \begin{equation}
            \textbf{GYRO}(\mathrm s)=\frac{200\pi}{\mathrm{s}+200\pi}\mathbf{I}_3
        \end{equation}
    \item A star sensor modeled as a first-order dynamics with cut-off frequency $16\pi$ rad/s:
        \begin{equation}
            \textbf{SST}(\mathrm s)=\frac{16\pi}{\mathrm{s}+16\pi}\mathbf{I}_3
        \end{equation}
    \item A loop delay modeled as a 2$^{nd}$ order Pade approximation with time delay $T_d = 0.0625$ s:
        \begin{equation}
            \textbf{DELAY}(\mathrm s)=\frac{T_d^2\mathrm{s}^2-6T_d\mathrm{s}+12}{T_d^2\mathrm{s}^2+6T_d\mathrm{s}+12}\mathbf{I}_3
        \end{equation}
\end{itemize}

A gyro-stellar observer $\mathbf{O}(\mathrm{s})$ is used to filter the gyro and the star sensor measurements. Its state-space realization is given by:
\begin{equation}
    \left\{\begin{array}{c}
         \dot{\mathbf{x}}_O=\begin{bmatrix}
             -0.1131 & -1 \\
             0.003948 & 0
         \end{bmatrix}\mathbf{x}_O + \begin{bmatrix}
             0.1131 & 1 \\
             -0.00394 & 0
         \end{bmatrix}\begin{bmatrix}
             \bm\omega_m^\mathcal{SC} \\
             \bm\Theta_m^\mathcal{SC}
         \end{bmatrix} \\
         \begin{bmatrix}
             \hat{\bm\omega}^\mathcal{SC} \\
             \hat{\bm{\Theta}}^\mathcal{SC}
         \end{bmatrix}=\begin{bmatrix}
             1 & 0\\
             -0.1131 & -1
         \end{bmatrix}\mathbf{x}_O+\begin{bmatrix}
             0 & 0 \\0.1131 & 1
         \end{bmatrix}\begin{bmatrix}
             \bm\omega_m^\mathcal{SC} \\
             \bm\Theta_m^\mathcal{SC}
         \end{bmatrix}
    \end{array}
    \right.
\end{equation}

with $\mathbf{x}_O$ the observer state vector.

The closed-loop generalized plant $\mathbf{P}(\mathrm{s},\bm{\Delta}^\mathcal{SC},\textcolor{blue}{\bm{\Pi}^\mathcal{SC}},\mathbf{K}_\mathrm{ACS}(\mathrm{s}))$ used for the robust control synthesis is shown in Fig. \ref{fig:AOCS}. Note that for distributed optimization the block $\textcolor{blue}{\bm{\Pi}^\mathcal{SC}}$ has not to be taken into account.

The following weighting filters are used to normalize the inputs and outputs: $\mathbf{W}_\mathrm{ext}=\mathbf{T}_\mathrm{ext}$, $\mathbf{W}^\mathrm{SST}_n = \sqrt{\textbf{PSD}^\mathrm{SST}}$, $\mathbf{W}^\mathrm{GYRO}_n = \sqrt{\textbf{PSD}^\mathrm{GYRO}}$, $\mathbf{W}_\mathrm{APE}=(\mathrm{diag}(\textbf{APE}))^{-1}$, $\mathbf{W}_\mathrm{RPE}=(\mathrm{diag}(\textbf{RPE}))^{-1}\frac{\Delta t_\mathrm{RPE}\mathrm s (\Delta t_\mathrm{RPE} \mathrm s + \sqrt{12})}{\Delta t_\mathrm{RPE}^2\mathrm{s}^2+6\Delta t_\mathrm{RPE}\mathrm{s}+12}$ and $\mathbf{W}_S = \frac{1}{\gamma}\mathbf{I}_3$.

\begin{figure}[ht!]
    \centering
    \includegraphics[width=.7\columnwidth]{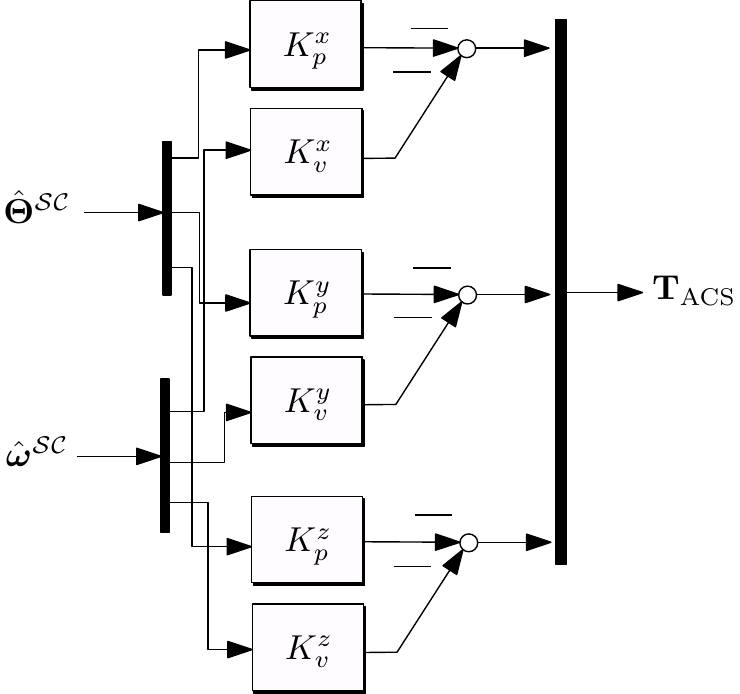}
    \caption{Structured controller $\mathbf{K}_\mathrm{ACS}(\mathrm{s})$}
    \label{fig:AOCS_block}
\end{figure}

Finally the block $\mathbf{K}_\mathrm{ACS}(\mathrm s)$ represents the structured $3\times 6$ attitude controller to be synthesized. The chosen structure is is shown in Fig. \ref{fig:AOCS_block}. It is a decentralized proportional-derivative controller (the  gains $K_p^i$  and $K_v^i$, $i=x,y,z$)

The set of the 6 controller tunable parameters ($K_p^i$, $w^i$  and $K_v^i$, $i=x,y,z$) are roughly initialized assuming a rigid 3-axis decoupled spacecraft and in order to:
\begin{itemize}
	\item reject a constant orbital disturbance $\mathbf{T}_\mathrm{ext}$  with a steady state pointing error $\bm{\Theta}^\mathrm{SC}$ lower than the $\textbf{APE}(i)$ requirement on each axis $i=x,y,z$,
	\item tune the 2$^{nd}$ order closed-loop dynamics of each axis with a damping ratio $\xi=0.7$  and a given frequency bandwidth $\omega^i$ ($\omega^i=0.06$ rad/s, $i=x,y,z$).
\end{itemize}

Indeed, under these assumptions, the open-loop model between the control torque and the pointing error is equal to $\frac{1}{\mathbf{J}_{B}^{\mathcal{SC}}s^2}$, where the nominal $3\times 3$ inertia $\mathbf{J}_{B}^{\mathcal{SC}}$ on the whole spacecraft at point $B$ can be computed from the DC gain of the nominal model $\left[\mathcal{M}_{B}^\mathcal{SC}\right]^{-1}_{\mathcal{R}_B}(\mathrm s)$:
\begin{equation}
	\mathbf{J}_{B}^{\mathcal{SC}} =\left[ \left[\mathcal{M}_{B}^\mathcal{SC}\right]^{-1}_{\mathcal{R}_B}(0)\right]^{-1}(4:6;4:6)
\end{equation}

Then the tuning:
\begin{align}
	\begin{split}
		K_{p}^i &=\mathbf J_B^{\mathcal{SC}} (i,i)\,\omega_i^{r^2} \\
		K_{v}^i &=2\,\xi \,\mathbf J_B^{\mathcal{SC}} (i,i)\,\omega_i^r
	\end{split}
 \label{eq:init_K}
\end{align}
with $\omega_i^r$ required bandwidth on $i$-th axis ($i=1,2,3$), ensures the needed closed-loop dynamics and a disturbance rejection function expressed as:
\begin{equation}
	\frac{\bm{\Theta}^\mathcal{SC} (i)}{\mathbf T_\mathrm{ext} (i)}({\mathrm{s}})=\frac{1}{\mathbf J_B^{\mathcal{SC}} (i,i)}\frac{1}{{{\mathrm{s}}}^2 +2\,\xi \,\omega_i^r {\mathrm{s}}+\omega_i^{r^2} }
\end{equation}
Thus the required bandwidth $\omega_i^r$ to meet the absolute pointing error requirement in steady state ($\bm{\Theta}^\mathcal{SC} (i) \le \textbf{APE}(i)$) is:
\begin{equation}
	\omega_i^r \ge \sqrt{\frac{\mathbf T_\mathrm{ext} (i)}{\mathbf J_B^{\mathcal{SC}} (i,i)\textbf{APE}(i)}}
\end{equation}

\begin{figure*}[t]
\begin{mdframed}[style=MyFrame]
\textbf{(I) Nested Control Optimization for Distributed Optimization Architecture:}
\begin{equation}
    J_{c_1}=\min\limits_{\mathbf{K}_\mathrm{ACS}}\max\limits_{\bm{\Delta}^\mathcal{SC}}\left\vert\left\vert\mathbf{P}_{
    \begin{bmatrix}
        \tilde{\mathbf{n}}^\mathrm{GYRO} \\
        \tilde{\mathbf{n}}^\mathrm{SST}
    \end{bmatrix}\rightarrow\begin{bmatrix}
        \tilde{\bm{\Theta}}^\mathcal{SC}_\mathrm{APE} \\
        \tilde{\bm{\Theta}}^\mathcal{SC}_\mathrm{RPE}
    \end{bmatrix}
    }(\mathrm{s},\bm{\Delta}^\mathcal{SC},\mathbf{K}_\mathrm{ACS})\right\vert\right\vert_2\le 1 
    \label{eq:objective_control_distributed}
\end{equation}

such that:
\begin{align}
    \begin{split}
       J_{c_2}=\max\limits_{\bm{\Delta}^\mathcal{SC}}\left\vert\left\vert\mathbf{P}_{
       \tilde{\mathbf{T}}_\mathrm{ext}\rightarrow\tilde{\bm{\Theta}}^\mathcal{SC}_\mathrm{APE}}
    (\mathrm{s},\bm{\Delta}^\mathcal{SC},\mathbf{K}_\mathrm{ACS})\right\vert\right\vert_\infty\le 1    \quad(\texttt{\textbf{Req1}})\\
     J_{c_3}=\max\limits_{\bm{\Delta}^\mathcal{SC}}\left\vert\left\vert\mathbf{P}_{
       \tilde{\mathbf{T}}_\mathrm{ext}\rightarrow\tilde{\bm{\Theta}}^\mathcal{SC}_\mathrm{RPE}}
    (\mathrm{s},\bm{\Delta}^\mathcal{SC},\mathbf{K}_\mathrm{ACS})\right\vert\right\vert_\infty\le 1    \quad(\texttt{\textbf{Req2}})\\
     J_{c_4}=\max\limits_{\bm{\Delta}^\mathcal{SC}}\left\vert\left\vert\mathbf{P}_{
       \tilde{\mathbf{T}}_\mathrm{ext}\rightarrow\tilde{\mathbf{u}}}
    (\mathrm{s},\bm{\Delta}^\mathcal{SC},\mathbf{K}_\mathrm{ACS})\right\vert\right\vert_\infty\le 1    \quad(\texttt{\textbf{Req3}})\\
    J_{c_5}=\max\limits_{\bm{\Delta}^\mathcal{SC}}\left\vert\left\vert\mathbf{P}_{
       \tilde{\mathbf{T}}_\mathrm{ext}\rightarrow\tilde{\mathbf{T}}}
    (\mathrm{s},\bm{\Delta}^\mathcal{SC},\mathbf{K}_\mathrm{ACS})\right\vert\right\vert_\infty\le 1    \quad(\texttt{\textbf{Req4}})\\
    \end{split}
    \label{eq:constraints_control_distributed}
\end{align}

\textbf{(II) Control/Structure Co-Optimization for Monolithic Optimization Architecture:}

\begin{equation}
    \left\{\begin{array}{c}
\min\limits_{\mathbf{K}_\mathrm{ACS},\textcolor{blue}{\bm{\Pi}^\mathcal{SC}}}\max\limits_{\bm{\Delta}^\mathcal{SC}}\left\vert\left\vert\mathbf{P}_{
    \begin{bmatrix}
        \tilde{\mathbf{n}}^\mathrm{GYRO} \\
        \tilde{\mathbf{n}}^\mathrm{SST}
    \end{bmatrix}\rightarrow\begin{bmatrix}
        \tilde{\bm{\Theta}}^\mathcal{SC}_\mathrm{APE} \\
        \tilde{\bm{\Theta}}^\mathcal{SC}_\mathrm{RPE}
    \end{bmatrix}
    }(\mathrm{s},\bm{\Delta}^\mathcal{SC},\textcolor{blue}{\bm{\Pi}^\mathcal{SC}},\mathbf{K}_\mathrm{ACS})\right\vert\right\vert_2\le 1   \\
      \min\limits_{\mathbf{K}_\mathrm{ACS},\textcolor{blue}{\bm{\Pi}^\mathcal{SC}}}\max\limits_{\bm{\Delta}^\mathcal{SC}}\Bar{\sigma}\left(\left[\left[\mathcal{M}_B^\mathcal{SC}\right]^{-1}_{{\mathbf{W}_{\mathrm{ext/\mathcal{B},B}}(1)\rightarrow \ddot{\mathbf{x}}_B(1)}
    }(\mathrm{j}\omega,\bm{\Delta}^\mathcal{SC},\textcolor{blue}{\bm{\Pi}^\mathcal{SC}})\right]^{-1}\right), \quad \forall \omega\in\left[0,\,0.0001\right]
    \end{array}
    \right.
    \label{eq:monolithic_control}
\end{equation}

such that:
\begin{align}
    \begin{split}
       \max\limits_{\bm{\Delta}^\mathcal{SC}}\left\vert\left\vert\mathbf{P}_{
       \tilde{\mathbf{T}}_\mathrm{ext}\rightarrow\tilde{\bm{\Theta}}^\mathcal{SC}_\mathrm{APE}}
    (\mathrm{s},\bm{\Delta}^\mathcal{SC},\textcolor{blue}{\bm{\Pi}^\mathcal{SC}},\mathbf{K}_\mathrm{ACS})\right\vert\right\vert_\infty\le 1    \quad(\texttt{\textbf{Req1}})\\
     \max\limits_{\bm{\Delta}^\mathcal{SC}}\left\vert\left\vert\mathbf{P}_{
       \tilde{\mathbf{T}}_\mathrm{ext}\rightarrow\tilde{\bm{\Theta}}^\mathcal{SC}_\mathrm{RPE}}
    (\mathrm{s},\bm{\Delta}^\mathcal{SC},\textcolor{blue}{\bm{\Pi}^\mathcal{SC}},\mathbf{K}_\mathrm{ACS})\right\vert\right\vert_\infty\le 1    \quad(\texttt{\textbf{Req2}})\\
     \max\limits_{\bm{\Delta}^\mathcal{SC}}\left\vert\left\vert\mathbf{P}_{
       \tilde{\mathbf{T}}_\mathrm{ext}\rightarrow\tilde{\mathbf{u}}}
    (\mathrm{s},\bm{\Delta}^\mathcal{SC},\textcolor{blue}{\bm{\Pi}^\mathcal{SC}},\mathbf{K}_\mathrm{ACS})\right\vert\right\vert_\infty\le 1    \quad(\texttt{\textbf{Req3}})\\
    \max\limits_{\bm{\Delta}^\mathcal{SC}}\left\vert\left\vert\mathbf{P}_{
       \tilde{\mathbf{T}}_\mathrm{ext}\rightarrow\tilde{\mathbf{T}}}
    (\mathrm{s},\bm{\Delta}^\mathcal{SC},\textcolor{blue}{\bm{\Pi}^\mathcal{SC}},\mathbf{K}_\mathrm{ACS})\right\vert\right\vert_\infty\le 1    \quad(\texttt{\textbf{Req4}})\\
    \end{split}
    \label{eq:hard_constraints_monolithic}
\end{align}
\end{mdframed}
\end{figure*}

This initial tuning, based on simplified assumptions is useful to initialize the non-convex optimization control problem. A robust controller is in fact synthesized thanks to the \texttt{systune} routine available in MATLAB, based on the non-convex optimization algorithm proposed by \cite{apkarian2015parametric}.

For the control synthesis two cases have to be distinguished according to the type of optimization architecture.

\paragraph{Distributed optimization}
In the case of distributed optimization all structural optimization parameters are fixed for each control synthesis. In that case the nested mixed $\mathcal{H}_2/\mathcal{H}_\infty$ control optimization problem is directly formulated from the objective (Eq. \eqref{eq:objective_control_distributed}) and the constraints (Eq. \eqref{eq:constraints_control_distributed}) defined in Section \ref{sec:robust_control}. 
The objective is in fact to minimize the $\mathcal{H}_2$-norm (or the variance) of the transfer from all measurements noises to the pointing requirements for the worst-case uncertainty configuration by coping with a set of hard constraints, expressed in terms of $\mathcal{H}_\infty$-norm. 

\paragraph{Monolithic optimization}
In the case a monolithic optimization architecture is chosen, the non-convex control synthesis algorithm can handle both control and structure optimization at the same time. In this case the hard constraints \eqref{eq:hard_constraints_monolithic}  are exactly the same as in the distributed optimization. However the multi-objectives problem shown in Eq. \eqref{eq:monolithic_control} (where $\Bar{\sigma}(\bullet)$ represents the upper bound singular value) has to be solved instead.

Note that the second objective (where the notation of the projections in body frame has been omitted for better clarity), translates the minimization of the overall spacecraft nominal mass.
The transfer function $\left[\left[\mathcal{M}_B^\mathcal{SC}\right]^{-1}_{\mathbf{W}_{\mathrm{ext/\mathcal{B},B}}(1)\rightarrow \ddot{\mathbf{x}}_B(1)}(\mathrm{j}\omega,\bm{\Delta}^\mathcal{SC},\textcolor{blue}{\bm{\Pi}^\mathcal{SC}})\right]^{-1}$ for $\omega\in\left[0,\,0.0001\right]$ represents in fact the DC gain (or the low frequency response) of the inverted transfer from the force to acceleration along the $x$-axis, that is the nominal mass of the overall spacecraft.

\section{Optimization methods}
\label{sec:optimization}

In this section the distributed and monolithic optimization algorithms used in this study are detailed.

\paragraph{Distributed architecture}

\begin{figure}[h!]
    \centering
    \includegraphics[width=\columnwidth]{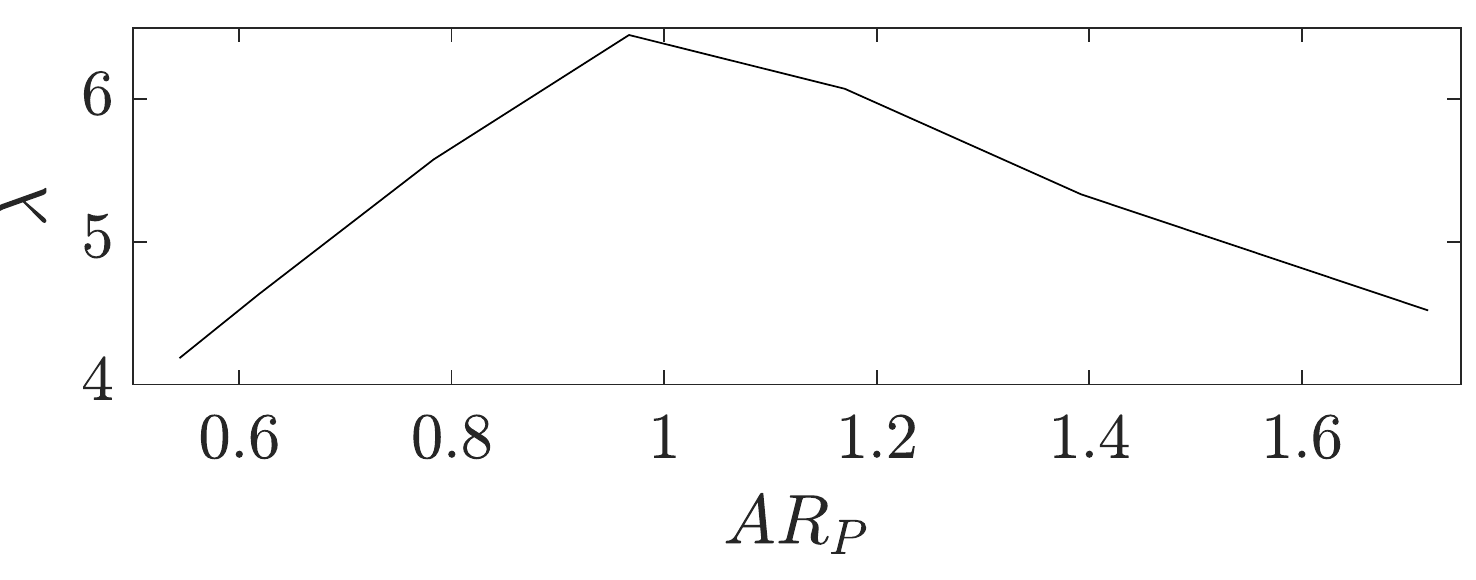}
    \caption{Launch constraint: computation of $\lambda$ factor}
    \label{fig:launch_constraint_lambda}
\end{figure}

In a distributed optimization architecture, a global optimization algorithm generates at each iteration $i$ a set of $N_s$ sub-iterations and corresponding $N_s$ vectors of structural optimization parameters $\bm\chi^{i,s}$, with $s=1,\dots,N_s$. At each sub-iteration $s$, a nested control optimization is solved as shown in Eq. \eqref{eq:nested}, where here $\mathbf{y}_s = {\bm\chi^{i,s}}$. The population $\bm\chi^{i+1,s}$ (with $s=1,\dots,N_s$) of the next iteration is then chosen in the imposed ranges $\mathbf{r}_\chi$ of each optimization parameter with a ratio based on the evaluation of the $N_s$ objective functions $J^s$ obtained at the previous iteration. The global optimization algorithm finally provides the best particle $\hat{\bm\chi}$ after some stopping criteria are reached. In this study the chosen global optimization algorithm is the Particle Swarm Optimization (PSO) (\cite{kennedy95}), implemented in the MATLAB Global Optimization Toolbox. 
The full distributed optimization routine proposed in this study is synthesized in Algorithm \ref{alg:distr}. 
Note that for each sub-iteration a child BDF NASTRAN file ($\mathrm{BDF\_C}$) is written from a parent one ($\mathrm{BDF\_P}$) for each flexible sub-structure $\mathcal{A}_j$. NASTRAN is then called from MATLAB to analyse the sub-structure and produce the corresponding f06 result file, that is then used in SDTlib to build the corresponding TITOP model $\left[\mathcal{M}_{A_j}^{\mathcal A_j}\right]_{\mathcal{R}_{A_j}}(\mathrm{s},\bm\Delta_{\mathcal A_j})$. Before proceeding with the nested optimization, a structural constraint has to be verified for the ENVISION benchmark. For structural safety of the solar panel in stowed configuration during the launch, the following empirical hard constraint is introduced:
\begin{equation}
    \omega_\mathrm{STO} = \lambda\sqrt{\frac{R_P E_Ph_{s_P}^3}{12\beta(1-\nu_P)^2}}\quad \left[\mathrm{rad/s}\right]
\end{equation}
where $R_P$ is the panel's second moment of area, $E_P$ is the skin Young's Modulus, $h_{s_P}$ is the total honeycomb skin thickness and $\nu_P$ is the Poisson's ratio. Note that the expression to compute the second moment of area is;
\begin{equation}
    R_P = \frac{12I_P}{h_{s_P}^3} 
\end{equation}
where $I_P=\left(\frac{t_{c_P}}{2}+\frac{h_{s_P}}{4}\right)^2$. The parameter $\beta$ is computed as:
\begin{equation}
    \beta = \frac{\rho_s+\rho_c}{l_P w_P}
\end{equation}
where $\rho_s$ and $\rho_c$ are respectively the mass per unit of area of the panel's skin and the core, $l_P$ is the length and $w_P$ the width of each panel.
Finally parameter $\lambda$ is a function of the panel aspect ratio $AR_P=l_P/w_P$ as shown in Fig. \ref{fig:launch_constraint_lambda}. 
To conclude, $\omega_\mathrm{STO}$ is a function of three optimization parameters ($t_{s_P}$, $t_{c_P}$ and $AR_P$), the other ones being constant.

The constraint to be satisfied is then $\omega_\mathrm{STO}>\omega_L$, where $\omega_L = 76\pi\,\mathrm{rad/s}$. In case this test is not passed then a large value is assigned to the objective function of the current swarm sub-iteration: $J^s=10$, in order to penalize the selection of this solution.
If the launch constraint is satisfied, then the TITOP model of the entire spacecraft is computed with SDTlib and the nominal mass $m$ can be recovered.
With the structural optimization parameters fixed, a control optimization is now possible by solving problem \eqref{eq:objective_control_distributed} constrained by \eqref{eq:constraints_control_distributed} with \textit{systune}. The global optimization optimization function of the current swarm sub-iteration can then be computed as:
\begin{equation}
    J^s = \frac{m}{\Bar{m}}+\sum_{j=1}^5 J_{c_j}
\end{equation}
where $\Bar{m}$ is the maximum expected spacecraft overall mass, obtained by imposing the maximum value of each structural parameter impacting the mass.
Finally the PSO will terminate when a maximum number of iterations $N_i$ is reached.

\begin{algorithm}
\caption{Distributed Optimization}\label{alg:distr}
\begin{algorithmic}[1]
\Inputs{$\mathbf{r}_\chi,\mathrm{BDF\_P},\omega_L,N_i,N_s,\bm{\Delta}^{\mathcal{SC}}$}
\Initialize{$i$, $\hat{J}$}
\While{$i \neq N_i$}
    \Create{$\bm\chi^{i,s}$, with $s=1,\dots,N_s$}
    \ForEach{$s$}
        \ForEach{flexible structure $\mathcal{A}_j$}
            \Create{$\mathrm{BDF\_C}$}
            \Run{NASTRAN analysis with $\bm\chi^{i,s}$}
            \Compute{$\left[\mathcal{M}_{A_j}^{\mathcal A_j}\right]_{\mathcal{R}_{A_j}}(\mathrm{s},\bm\Delta_{\mathcal A_j})$} 
        \EndFor
        \Compute{$\omega^{s}_\mathrm{STO}$} 
        \If{$\omega^{s}_\mathrm{STO}\le\omega_L$}
            \State $J^s\gets 10$
            \ElsIf {$\omega^{s}_\mathrm{STO}>\omega_L$}
                \Compute{$\left[\mathcal{M}_{B}^{\mathcal{SC}}\right]^{-1}_{\mathcal{R}_B}(\mathrm{s},\bm\Delta^{\mathcal{SC}})$} 
                \State $m\gets \left[\mathcal{M}_B^\mathcal{SC}\right]^{-1}_{\mathbf{W}_{\mathrm{ext/\mathcal{B},B}}(1)\rightarrow \ddot{\mathbf{x}}_B(1)}(0)$
                \Initialize{$\mathbf{K}_\mathrm{ACS}$} \Comment{See Eq. \eqref{eq:init_K}}
                \Solve{\eqref{eq:objective_control_distributed} subject to \eqref{eq:constraints_control_distributed}}
                \State $J^s \gets \frac{m}{\Bar{m}}+\sum_{j=1}^5 J_{c_j}$
                    \If{$J^s<\hat{J}$}
                        \State $\hat{J}\gets J^s$
                        \State $\hat{\bm\chi}\gets \bm\chi^{i,s}$
                        \State $\hat{\mathbf{K}}_\mathrm{ACS}\gets \hat{\mathbf{K}}^s_\mathrm{ACS}$
                    \EndIf
        \EndIf
        \State $i\gets i+1$
    \EndFor
\EndWhile
\State \Return $\hat{\bm\chi}$, $\hat{\mathbf{K}}_\mathrm{ACS}$
\end{algorithmic}
\end{algorithm}

\paragraph{Monolithic architecture}

In the monolithic optimization architecture, the control optimization routine handles the structure optimization as well as already shown in Section \ref{sec:robust_control}. Algorithm \ref{alg:mono} lists the main stpdf.

\begin{algorithm}
\caption{Monolithic Optimization}\label{alg:mono}
\begin{algorithmic}[1]
\Inputs{$\mathbf{r}_\chi,\bm{\Delta}^{\mathcal{SC}}$}
\Initialize{$\bm\Pi^\mathcal{SC}$} \Comment{See Section \ref{sec:delta_monomithic}}
\Compute{$\left[\mathcal{M}_{B}^{\mathcal{SC}}\right]^{-1}_{\mathcal{R}_B}(\mathrm{s},\bm\Delta^{\mathcal{SC}},\bm\Pi^{\mathcal{SC}})$} 
\Initialize{$\mathbf{K}_\mathrm{ACS}$} \Comment{See Eq. \eqref{eq:init_K}}
\Solve{\eqref{eq:monolithic_control} subject to \eqref{eq:hard_constraints_monolithic}}
\Compute{$\hat{\bm\chi}$ from $\hat{\bm\Pi}^\mathcal{SC}$}
\State \Return $\hat{\bm\chi}$, $\hat{\mathbf{K}}_\mathrm{ACS}$
\end{algorithmic}
\end{algorithm}

\section{Results}
\label{sec:results}

\begin{figure}[!ht]
    \centering
    \includegraphics[width=\linewidth]{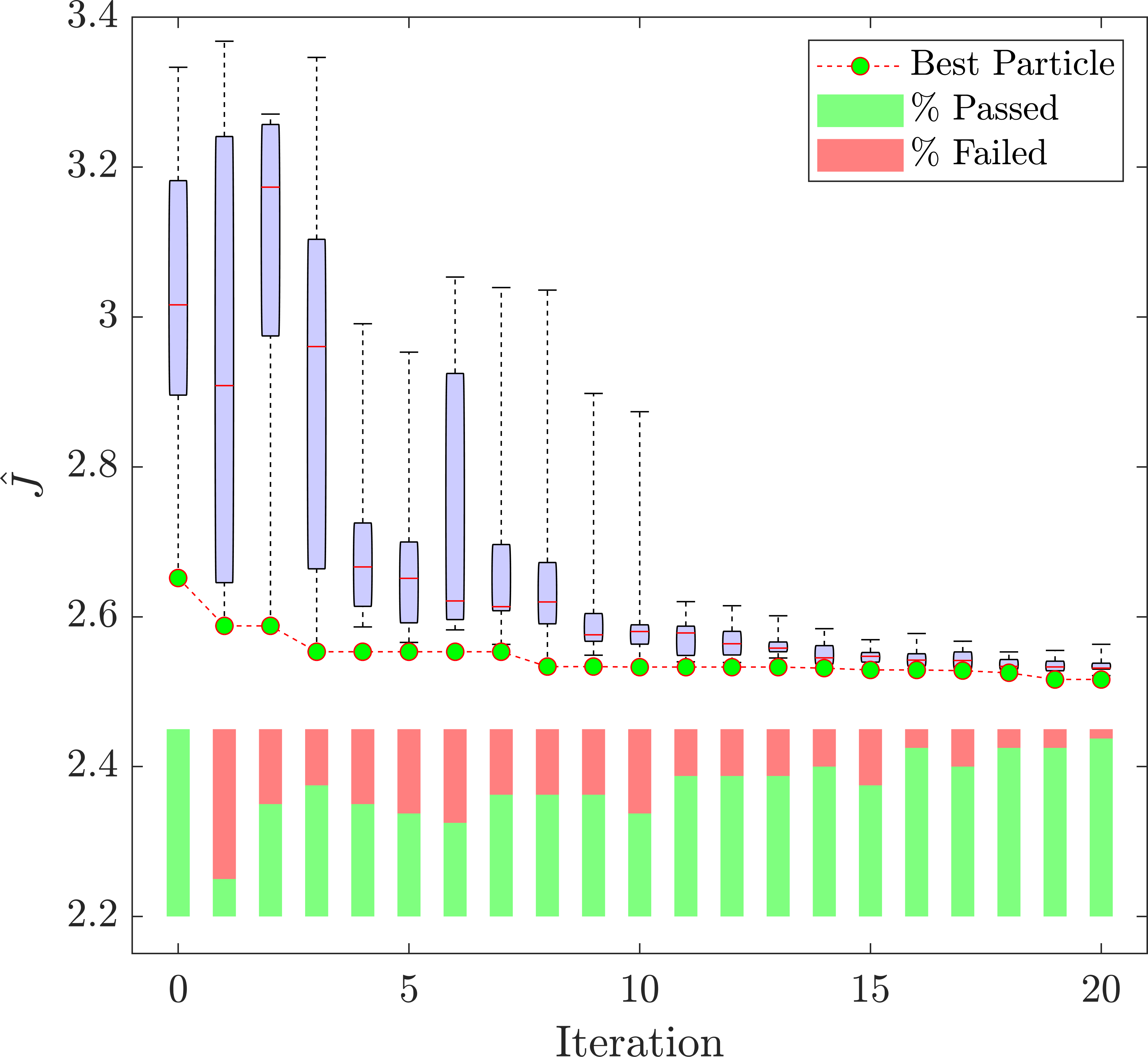}
    \caption{Evolution of the objective function along the PSO iterations}
    \label{fig:fBest}
\end{figure}

For the distributed optimization of the ENVISION benchmark the 12 structural optimization parameters in Table \ref{tab:opt_param} are considered. After a trial and error process the maximum number of PSO iterations and sub-iterations have been fixed to $N_i=20$ and $N_s=20$ respectively as compromise between computational time and convergence to the optimal solution. Figure \ref{fig:fBest} shows the evolution of the objective function $\hat{J}$ along the iterations, with quantiles (blue boxes) of the swarm particles having passed the launch constraint test. The dispersion of the particles reduces along the iterations by giving an indication on the convergence to the optimal solution.

\begin{table*}[ht!]
    \centering
    \resizebox{\textwidth}{!}{\begin{tabular}{lcccccc}
    \hline
    \textbf{Parameter} & \textbf{Symbol} & \textbf{Unit} & \textbf{Min Value} & \textbf{Max Value} & \textbf{Distributed Opt} & \textbf{Monolithic Opt} \\ \hline
    Yoke Young Modulus & $E_Y$ & $\mathrm{Pa}$ & $1.1\cdot 10^{11}$ & $1.23\cdot 10^{11}$ & $1.121\cdot 10^{11}$ &  ($1.165\cdot 10^{11}$)\\
    Yoke Density & $\rho_Y$ & $\mathrm{kg/m^3}$ & $2.18\cdot 10^3$ & $4.5\cdot 10^3$ & $4.186\cdot 10^3$ & ($3.340\cdot 10^3$)\\
    Yoke Section Length & $B_Y$ & $\mathrm{m}$ & $1.5\cdot 10^{-2}$ & $5\cdot 10^{-2}$ & $1.544\cdot 10^{-2}$ & ($3.25\cdot 10^{-2}$)\\
    Yoke Section Height  & $D_Y$ & $\mathrm{m}$ & $1.5\cdot 10^{-2}$ & $5\cdot 10^{-2}$ & $1.502\cdot 10^{-2}$ & ($3.25\cdot 10^{-2}$)\\
    Yoke Section Thickness & $t_Y$ & $\mathrm{m}$ & $1\cdot 10^{-3}$ & $2\cdot 10^{-3}$ & $1.017\cdot 10^{-3}$ & ($1.5\cdot 10^{-3}$)\\
    Panel Skin Thickness & $t_{s_P}$ & $\mathrm{m}$ & $2\cdot 10^{-4}$ & $4\cdot 10^{-4}$ & $2.042\cdot 10^{-4}$ & $2\cdot 10^{-4}$  \\
    Panel Core Thickness & $t_{c_P}$ & $\mathrm{m}$ & $1\cdot 10^{-2}$ & $3.5\cdot 10^{-2}$ & $1.322\cdot 10^{-2}$ & $1\cdot 10^{-2}$\\
    Yoke Length Ratio & $LR_Y$ & $\mathrm{-}$ & $0.42$ & $1$ & $1$ & ($0.71$) \\
    Panel Aspect Ratio & $AR_P$ & $\mathrm{-}$ & $3/4$ & $4/3$ & $1.237$ & ($25/24$) \\
    SRS Section Outer Radius & $R_\mathrm{SRS}$ & $\mathrm{m}$ & $1.25\cdot 10^{-2}$ & $2\cdot 10^{-2}$ & $1.25\cdot 10^{-2}$ & $1.25\cdot 10^{-2}$\\
    SRS Section Thickness & $t_{\mathrm{SRS}}$ & $\mathrm{m}$ & $3.8\cdot 10^{-4}$ & $6\cdot 10^{-4}$ & $4.959\cdot 10^{-4}$ & ($4.9\cdot 10^{-4}$)\\
    SAR Core Thickness & $t_{c_V}$ & $\mathrm{m}$ & $5\cdot 10^{-4}$& $1.5\cdot 10^{-3}$ & $5.178\cdot 10^{-4}$ & $1.5\cdot 10^{-3}$\\
        \hline
    \end{tabular}}
    \caption{Optimization structural parameters. Note: for the monolithic optimization the values in brackets correspond to the set of fixed parameters}
    \label{tab:opt_param}
\end{table*}

\begin{figure*}[!ht]
    \centering
    \includegraphics[width=\linewidth]{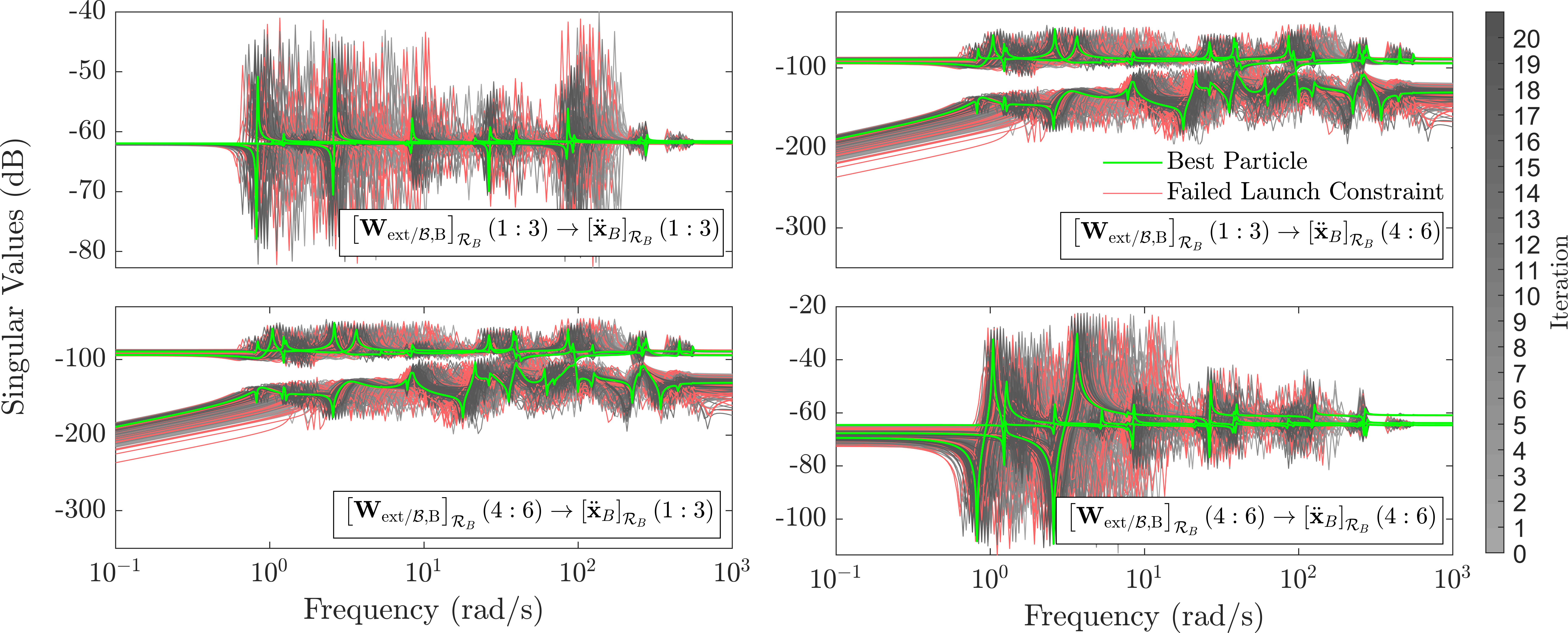}
    \caption{Singular values of all swarm particles along the optimization iterations}
    \label{fig:sigma_iter}
\end{figure*}

Figure \ref{fig:sigma_iter} shows the evolution of the singular values for all particle swarm iterations. Notice as the optimal solution attracts all modes' resonances to lower frequencies by making the spacecraft lighter and more flexible.

\begin{figure*}[!ht]
    \centering
    \includegraphics[width=\linewidth]{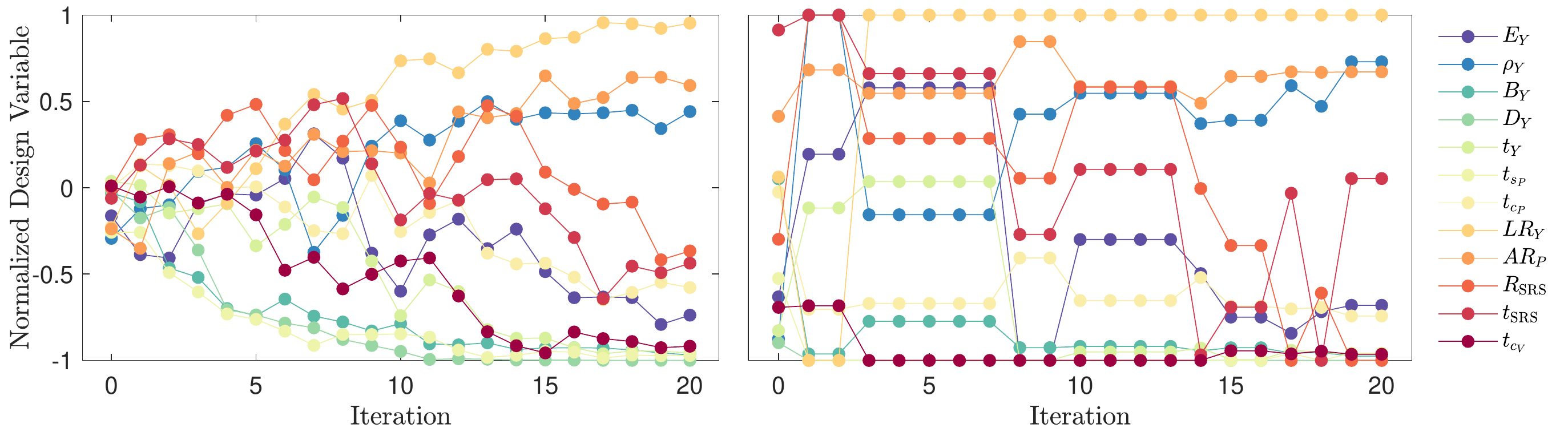}
    \caption{Evolution of the optimal structure parameters along the PSO iterations: average value among particle swarms of the same iteration (left) and best particle (right)}
    \label{fig:paramEvol}
\end{figure*}

A detail of the evolution of each optimization parameter is provided in Fig. \ref{fig:paramEvol}.

For the monolithic optimization, four optimization parameters are chosen ($t_{s_P}$, $t_{c_P}$, $R_\mathrm{SRS}$ and $t_{c_V}$) as already mentioned in Section \ref{sec:delta_monomithic}. All the other parameters are fixed to their mean value in order to respect the launch constraint. This constraint cannot in fact easily be included in the monolithic approach as in the distributed one.

The optimal mechanical parameters obtained with both distributed and monolithic optimization are shown in Table \ref{tab:opt_param}. We notice as parameters mostly impacting the mass (Fig. \ref{fig:MassSaving}) tend to reach their minimum value except for the SAR core thickness, that surprisingly reaches its maximum value in the monolithic optimization. This result can be interpreted with the fact that the monolithic optimization is based on a non-smooth gradient based optimization, that could fall into a local minimum. However, the advantage of using a monolithic optimization is the needed computational time with respect to the distributed architecture as shown in Table \ref{tab:general_perf}. The optimization total time for the distributed architecture is 12.33 hours versus $\approx$ 0.4 hours needed for the monolithic one. It has to be said that for the monolithic architecture a previous initialization of $\bm{\Pi}^\mathcal{SC}$ (step 1 in Algorithm \ref{alg:mono}) is not here taken into account. This time depends on the number of NASTRAN analyses to be run to have the initial set of samples to be interpolated with APRICOT. This operation is generally really fast and for the present application stayed smaller than $\approx 0.5$ hours. Finally, the APRICOT generation of $\bm{\Pi}^\mathcal{SC}$ took just $6.335$ s. Notice that both optimizations have been run on a Windows 7, 64 bits, Intel(R) Core(TM) i7-4810MQ CPU @ 2.80GHz, RAM 16 Go. Even if the achieved optimized mass is similar with the two approaches,
if with a distributed architecture a big number of optimization parameters can be considered (with a
consequent lower computational time), with the monolithic optimization a small set of mechanical
parameters can be used. This number is in fact limited as seen in the previous sections by the complexity of the interpolated LFT, that can contain a huge number of uncertain repeated parameters, which can make the robust control optimization infeasible. In this study four parameters out of twelve have been chosen based on their impact on the overall spacecraft mass. This choice on the other hand constraints the achievable control performance, that is driven by other parameters as discussed in Section \ref{sec:analysis_distributed}.

\begin{table}[!ht]
    \centering
    \resizebox{\columnwidth}{!}{\begin{tabular}{lrr}
    \hline
        & \textbf{Distributed} & \textbf{Monolithic} \\ \hline
        Opt.Total Time (h) & 12.33 &  0.4072 \\
        Opt. SC Total Mass (kg) & $1253.86$ & $1258.166$\\ 
        Total Mass reduction (kg) & $55.94$ & $51.63$\\
        $\%$ Opt. Total Mass & $4.27\%$ & $3.94$\%\\
        Opt. Max Control Perf. & $0.7208$ & $1.0011$\\ \hline
    \end{tabular}}
    \caption{General performance of distributed and monolithic optimization}
    \label{tab:general_perf}
\end{table}

As shown in Table \ref{tab:general_perf}, the achieved mass reduction is almost the same with the two approaches. Notice that the nominal mass of the central body corresponds to 1173 kg, that means that the actual percentage of saved mass with respect to the potential reducible mass (of the flexible appendages) corresponds indeed to $69.18\%$ and $63.85\%$, for the distributed and monolithic optimization respectively.  
Concerning the control performance with the distributed architecture, the maximum control index ($\max\limits_j {J_{c_j}}=0.7208<1$ with $j=1,\dots,5$) shows that control performance is largely satisfied. For monolithic architecture, this index is slightly greater than unity, while acceptable by keeping in mind that a large uncertainty level has been considered (see Table \ref{tab:uncertainty_list}). Details regarding each achieved control index are is provided in Table \ref{tab:contr_perf}. From this table, the APE and Sensitivity indexes result to be the most critical to be satisfied.

\begin{table}[!ht]
    \centering
    \resizebox{\columnwidth}{!}{\begin{tabular}{lrr}
    \hline
        \textbf{Cont. Req.} & \textbf{Distributed} & \textbf{Monolithic} \\ \hline
        \textbf{APE} & 0.7208 ($\bar{\mu}_\Delta$=0.7353) &  1.0011 ($\bar{\mu}_\Delta$=1.00051)\\
        \textbf{RPE} & 0.0583 & 0.1016\\ 
        \textbf{Command} & 0.0122 & 0.0157\\
        \textbf{Sensitivity} & 0.7208 ($\bar{\mu}_\Delta$=0.7530)& 1.0011 ($\bar{\mu}_\Delta$=1.01)\\
        \textbf{Noise Variance} & 0.0469 & 0.0521\\ \hline
    \end{tabular}}
    \caption{Optimized control performance with distributed and monolithic optimization}
    \label{tab:contr_perf}
\end{table}

Table \ref{tab:ACS_optimal} finally provides the optimal control gains obtained with both optimization approaches.

\begin{table}[h!]
    \centering
    \begin{tabular}{lcc}
    \hline
    & \textbf{Distributed} & \textbf{Monolithic} \\ \hline
    $K_p^x$ & 35.0764 (8.6952) & 35.2731 (8.8499)\\
    $K_v^x$ & 335.2577 (202.8248) & 201.3181 (206.4978)\\
    $K_p^y$ & 13.9779 (6.1029) & 14.8800 (6.2416)\\
    $K_v^y$ & 280.8861 (142.4001) & 162.0921 (145.6371)\\
    $K_p^z$ & 35.008 (10.5454) & 35.2562 (10.7226)\\
    $K_v^z$ & 404.4305 (246.0605) & 227.9570 (250.1946)\\ \hline
    \end{tabular}
    \caption{Optimal control parameters obtained with distributed and monolithic optimization. Values in brackets are the initial guess obtained with Eq. \eqref{eq:init_K}}
    \label{tab:ACS_optimal}
\end{table}

A deeper analysis is proposed in Section \ref{sec:analysis_distributed} to better interpret physically the achieved results by using the outcomes of the distributed optimization.

\subsection{Further analysis with distributed optimization}
\label{sec:analysis_distributed}

\begin{figure*}[!ht]
    \centering
    \includegraphics[width=.95\linewidth]{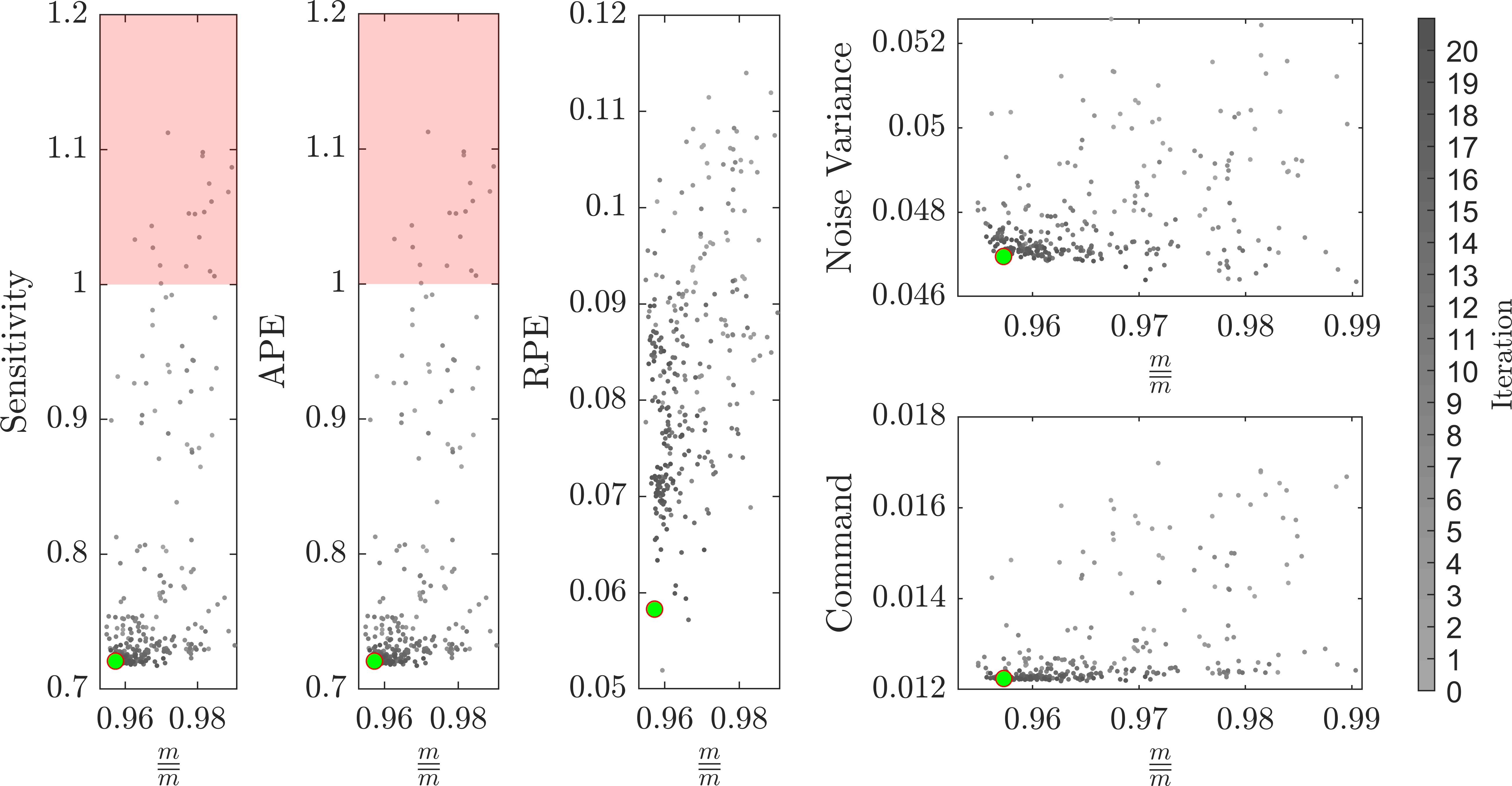}
    \caption{Pareto fronts for the distributed optimization: control versus structure performance indexes. The green bullet represents the best particle}
    \label{fig:paretoFrontBW}
\end{figure*}

The concurrent character of the proposed structure/control optimization is shown in the Pareto fronts of Fig. \ref{fig:paretoFrontBW}. In particular, a reduction of the spacecraft mass corresponds to a degradation of the pointing performance (APE channel) and stability (Sensitivity channel). The optimal solution (green bullet) represents the compromise between these competing objectives.

\begin{figure}[!ht]
    \centering
    \includegraphics[width=\linewidth]{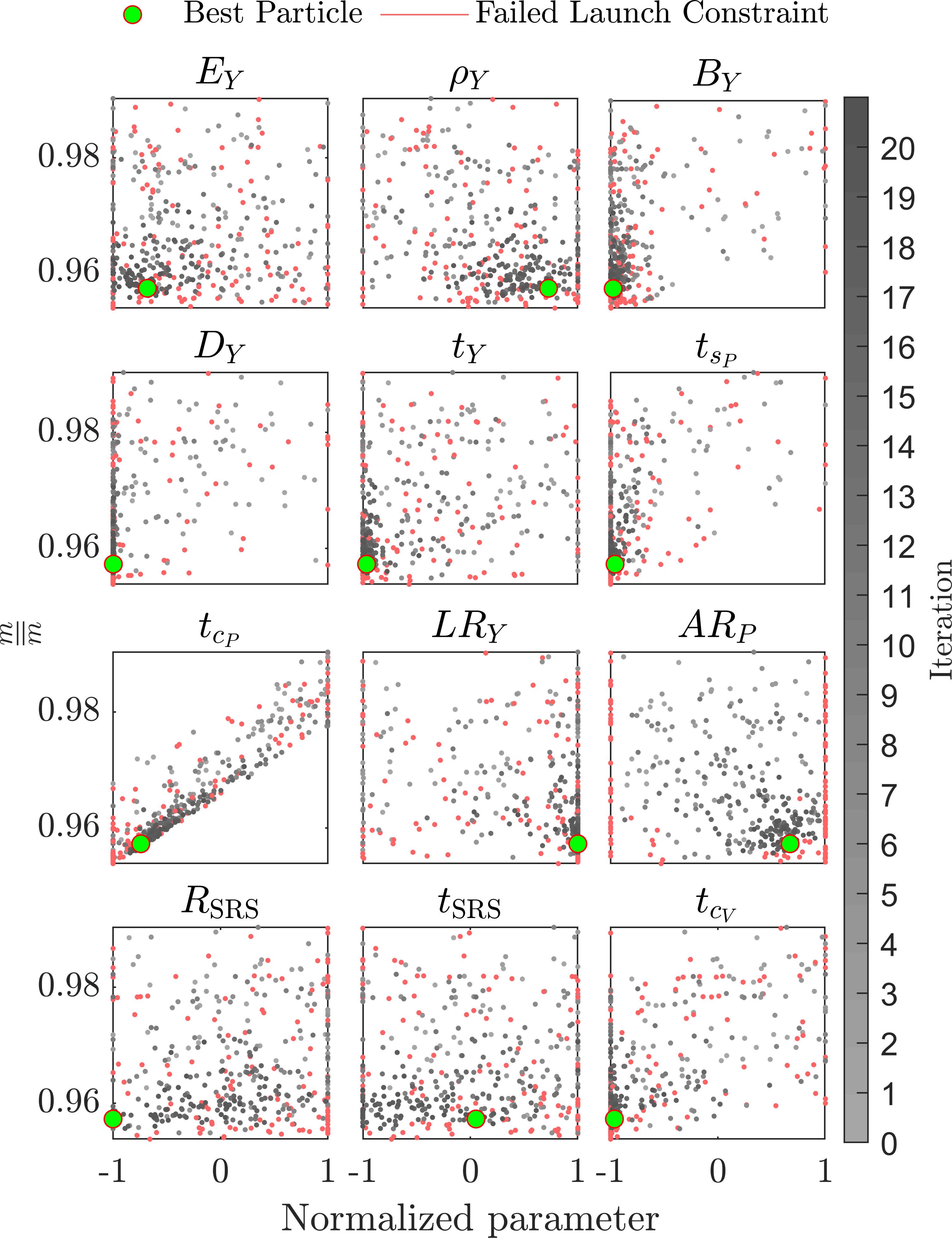}
    \caption{Mass optimization cost function versus structure optimization parameters}
    \label{fig:massVsparam}
\end{figure}

\begin{figure}[!ht]
    \centering
    \includegraphics[width=\linewidth]{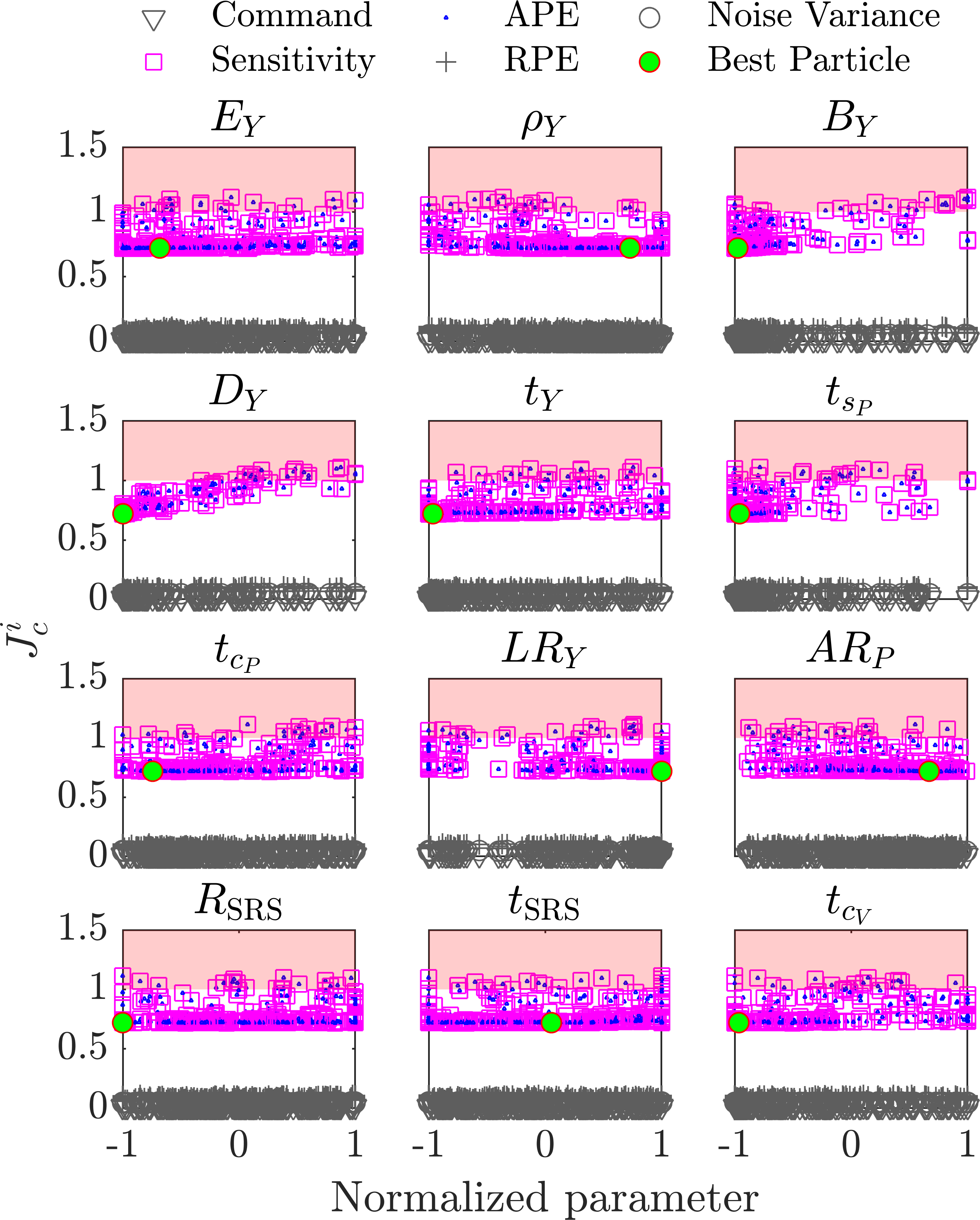}
    \caption{Control optimization cost function versus structure optimization parameters}
    \label{fig:ControlPerfVsparam}
\end{figure}

The impact of each structural optimization parameter on both structural and control optimization indexes can be highlighted by plotting the evolution of the particle swarms along the PSO iterations as done in Figs. \ref{fig:massVsparam} and \ref{fig:ControlPerfVsparam} respectively. The linear dispersion of particles according to a variation of $t_{c_P}$ confirms the importance of this parameter for a reduction of the spacecraft mass. A similar behavior, while less marked, can be noticed for $t_{s_P}$ and $t_{c_V}$, as expected.
Another observation is that a square-like configuration of the solar panels is preferred to a rectangular one since particles with extreme values of $AR_P$ are discarded since the launch constraint is not satisfied.

According to control performance, Fig. \ref{fig:ControlPerfVsparam} reveals the key structural parameters having the most important impact. As already shown in Table \ref{tab:contr_perf}, APE and Sensitivity indexes are the ones driving the overall control performance. They get the same value in the vast majority of cases according to the gradient-based algorithm implemented in the \textit{systune} routine. By looking at the dispersion of particles along the iterations, a clear dependence (linear) of control performance is noticed when Yoke's section dimensions (mostly $B_Y$ and $D_Y$) vary. In particular a degradation is experienced when bigger values of $B_Y$ and $D_Y$ are used.

\begin{figure}[!ht]
    \centering
    \includegraphics[width=\columnwidth]{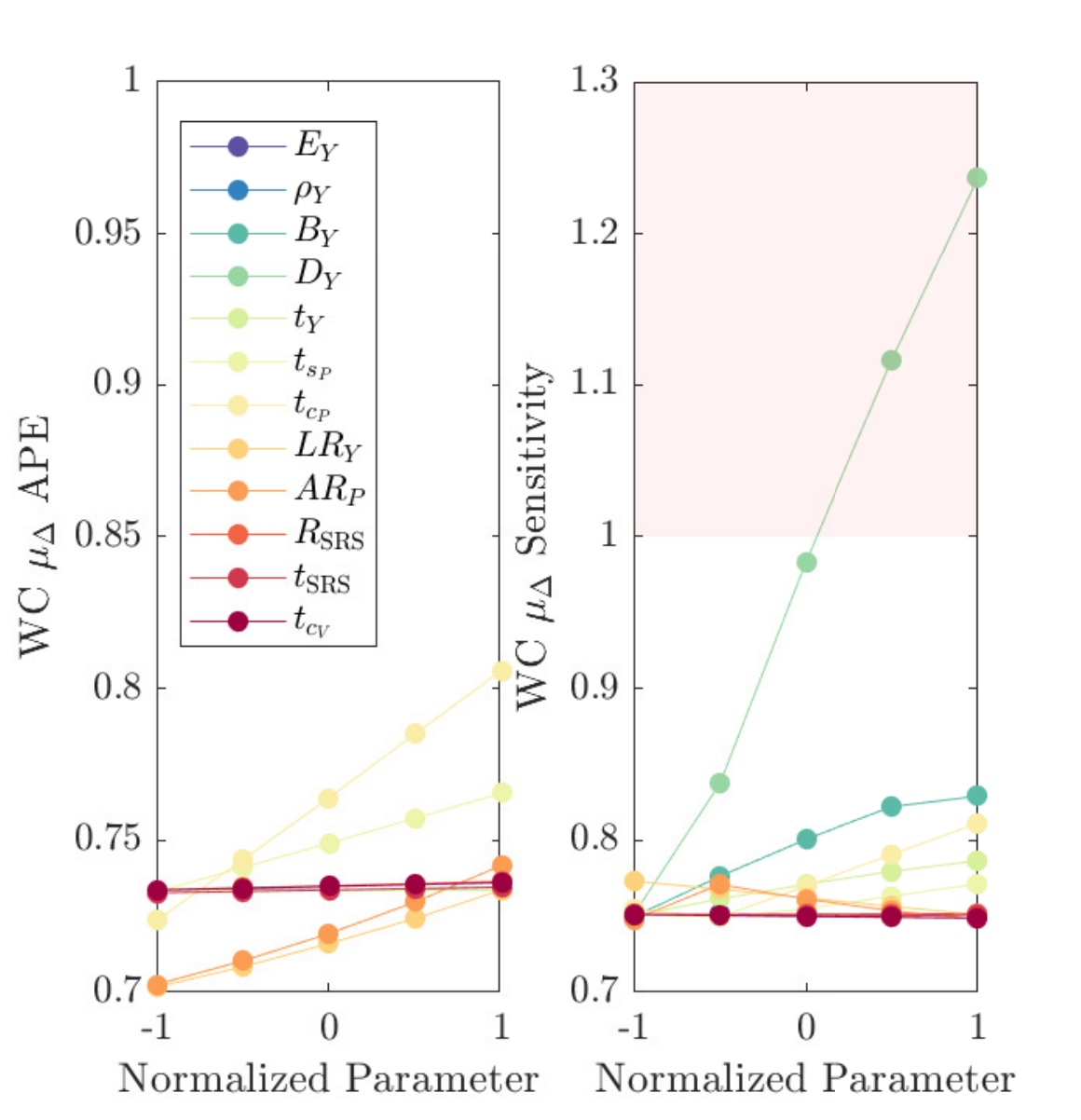}
    \caption{Sensitivity of the worst-case peak gain for APE and Sensitivity channel around the optimal solution}
    \label{fig:OptimalSolutionSensitivity}
\end{figure}

\begin{figure}[!ht]
    \centering
    \includegraphics[width=\columnwidth]{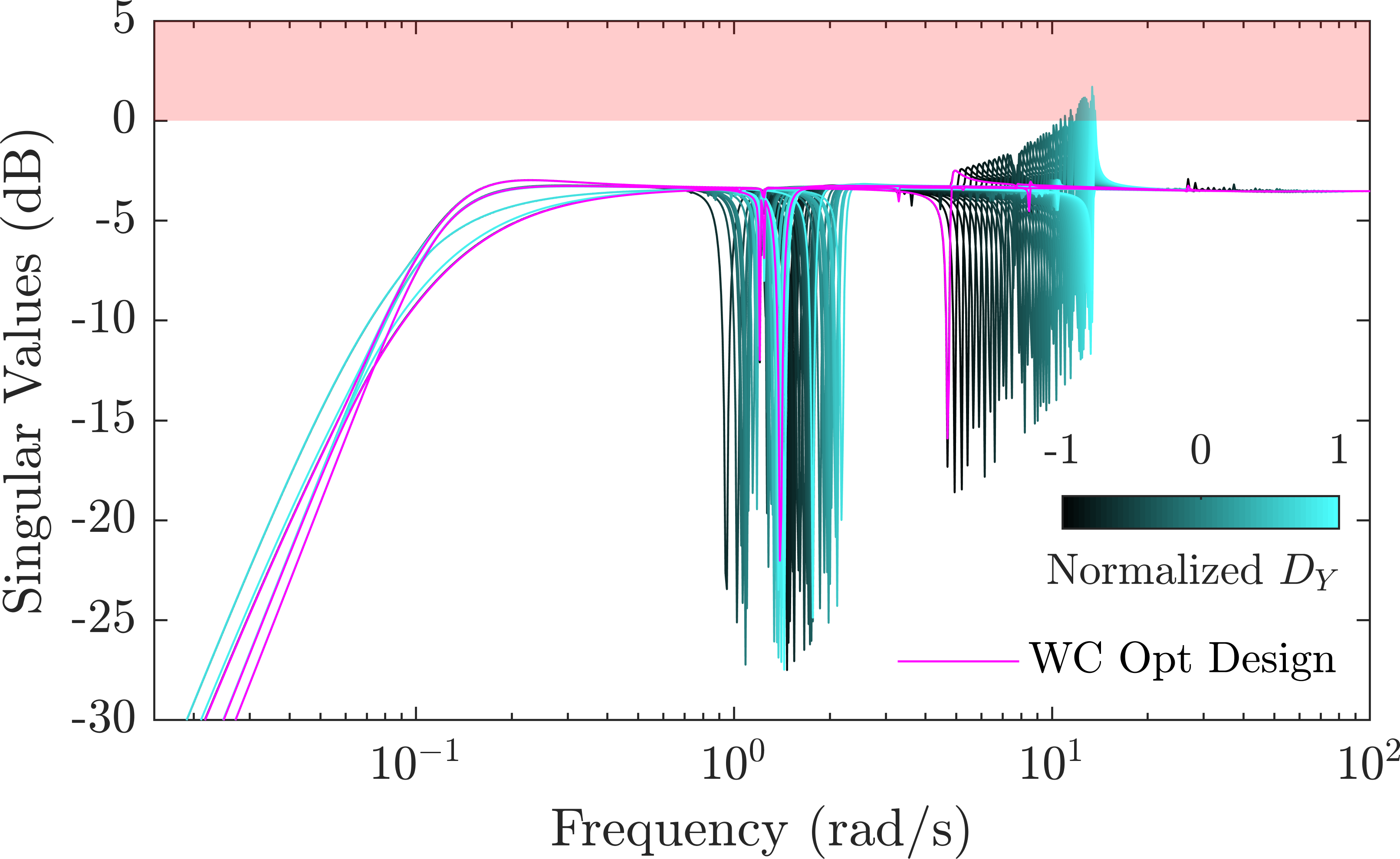}
    \caption{Estimation of WC Sensitivity control performance with variation of $D_Y$ in the optimal distributed design}
    \label{fig:Sensitivity_Dy}
\end{figure}

In order to better analyze this impact on the control performance, a further analysis is performed: worst-case (WC) APE and Sensitivity performance are checked when one optimization parameter is left varying by keeping all the others equal to the values corresponding to the optimal solution. In order to speed up the computation, a formal $\mu$-analysis is not applied for this analysis and \textit{systune} is used to obtain the worst-case peak gain of the demanded transfer functions (APE and Sensitivity channels). The research of the worst-case configuration in \textit{systune} is in fact based on a heuristic search and not a formal one. For this reason a formal validation of the synthesized controller through a $\mu$-analysis is proposed in Section \ref{sec:validation}. 
Since here the objective is to show the overall dependence of the control performance from the structural parameters around the optimal configuration an estimation of the WC is sufficient. The WC analysis is then turned into a parametric robust control design problem: the upper bound $\bar{p}_\mathrm{WC}$ of the peak gain $p_\mathrm{WC}$ of the analyzed transfer function is considered as a decision variable and the objective is to minimize $\bar{p}_\mathrm{WC}$ whilst meeting the constraint:
\begin{equation}
    \max\limits_{\bm\Delta^\mathcal{SC}\in\mathcal{D}_{\bm\Delta^\mathcal{SC}}}\left\vert\left\vert \frac{\mathbf{H}(\mathrm{s},\bm\Delta^\mathcal{SC})}{\bar{p}_\mathrm{WC}}
    \right\vert\right\vert_\infty\le 1
\end{equation}
where $\mathbf{H}(\mathrm{s},\bm\Delta^\mathcal{SC})$ is the transfer function to be checked and $\mathcal{D}_{\bm\Delta^\mathcal{SC}}$ is a subset of the parametric domain in which the parametric uncertainties $\bm\Delta^\mathcal{SC}$ can vary. This subset is chosen by \textit{systune} as said before.
By running this analysis, the results are provided in Fig. \ref{fig:OptimalSolutionSensitivity}. A first observation is that a variation of all parameters around the optimal solution does not critically affect the APE performance since the WC stays below unity. However, a degradation of the pointing performance is experienced with growing values of $t_{c_P}$, $t_{s_P}$, $LR_Y$ and $AR_P$. More critical is the impact of a variation of the parameters on the spacecraft stability (Sensitivity channel). An increase of one of the dimension of the Yoke's section ($D_Y$) around the optimal solution corresponds in fact to a fast degradation of the stability margins. This phenomenon is better highlighted in Fig. \ref{fig:Sensitivity_Dy}, where the singular values of the worst-case Sensitivity channel are plotted with a variation of $D_Y$ around the optimal solution. An amplification of a flexible mode is experienced with growing values of $D_Y$ and a maximum loss of $\approx 20\%$ of stability performance is reached.
This analysis explains also the worst control performance obtained with the monolithic approach since an average values is chosen for $D_Y$.

\subsection{Controller formal validation}
\label{sec:validation}

In order to certify the synthesized controllers both for distributed and monolithic optimization it is necessary to rigorously validate the worst-case control performance. As already mentioned, \textit{systune} routine is based on a heuristic search of the worst-case configurations. Formal validation with the computation of the structured singular value $\mu_\Delta$ is thus needed \cite{zhou1998essentials}. 
The uncertain parameter $\sigma_4$ related to the solar panel geometrical configuration is repeated 32 times in the uncertain block $\bm \Delta^\mathcal{SC}$ by leading to unacceptable computational time using the standard worst-case analysis tools. This problem is circumvented by sampling $\sigma_4$ on a grid of $N_\tau=50$ points regularly distributed in $[0,\;1]$. This subset has been chosen to account for the symmetric configuration of the model in the $\theta_\mathrm{SA} \in [0, 180]^\circ$ and $\theta_\mathrm{SA} \in [-180, 0]^\circ$ intervals. 

The computation of $\mu_\Delta$ lower and upper bounds is performed thanks to the \textit{wcgain} routine in MATLAB Robust Control Toolbox. 
The worst-case parametric configuration is associated to the lower bound $\underline\mu_\Delta$. The upper bound $\bar\mu_\Delta$ computation provides the conservatism in the estimation of the true value of $\mu_\Delta$.
Note that this analysis is only performed for the two most critical control performance, the APE and Sensitivity (see Table \ref{tab:contr_perf}):
\begin{align}
\begin{split}
\mu_\Delta^\mathrm{APE}&=\max\limits_{\bm{\Delta}^\mathcal{SC}}\left\vert\left\vert\mathbf{P}_{
       \tilde{\mathbf{T}}_\mathrm{ext}\rightarrow\tilde{\bm{\Theta}}^\mathcal{SC}_\mathrm{APE}}
    (\mathrm{s},\bm{\Delta}^\mathcal{SC},\mathbf{K}_\mathrm{ACS})\right\vert\right\vert_\infty \\
    \mu_\Delta^\mathrm{Sensitivity}&=\max\limits_{\bm{\Delta}^\mathcal{SC}}\left\vert\left\vert\mathbf{P}_{
       \tilde{\mathbf{T}}_\mathrm{ext}\rightarrow\tilde{\mathbf{T}}}
    (\mathrm{s},\bm{\Delta}^\mathcal{SC},\mathbf{K}_\mathrm{ACS})\right\vert\right\vert_\infty
    \end{split}
\end{align}

\begin{figure*}[!htb]
    \centering
    \includegraphics[width=\linewidth]{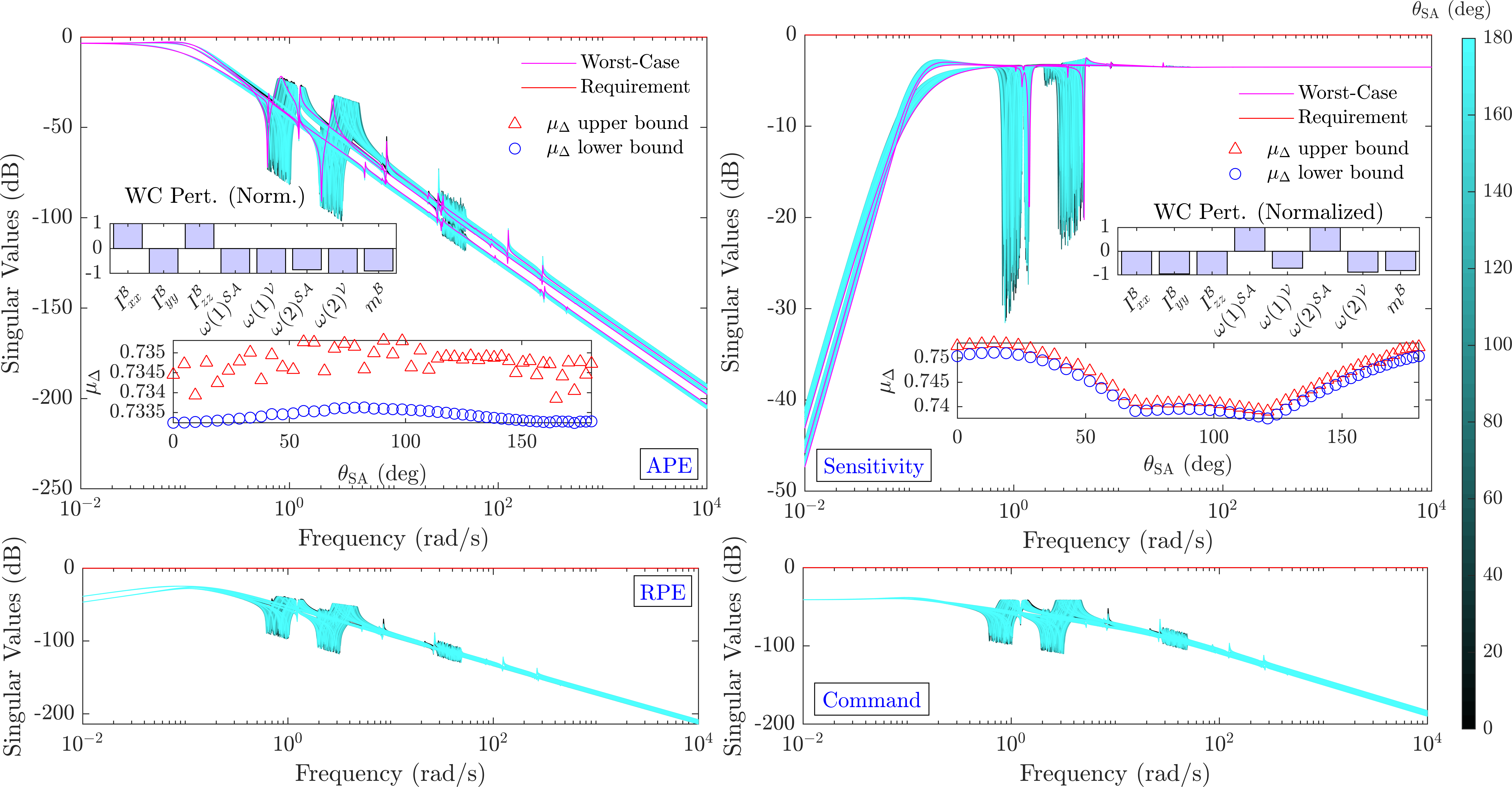}
    \caption{Worst-case analysis of control performance for distributed optimization}
    \label{fig:ControlPerf_Distributed}
\end{figure*}

\begin{figure*}[!htb]
    \centering
    \includegraphics[width=\linewidth]{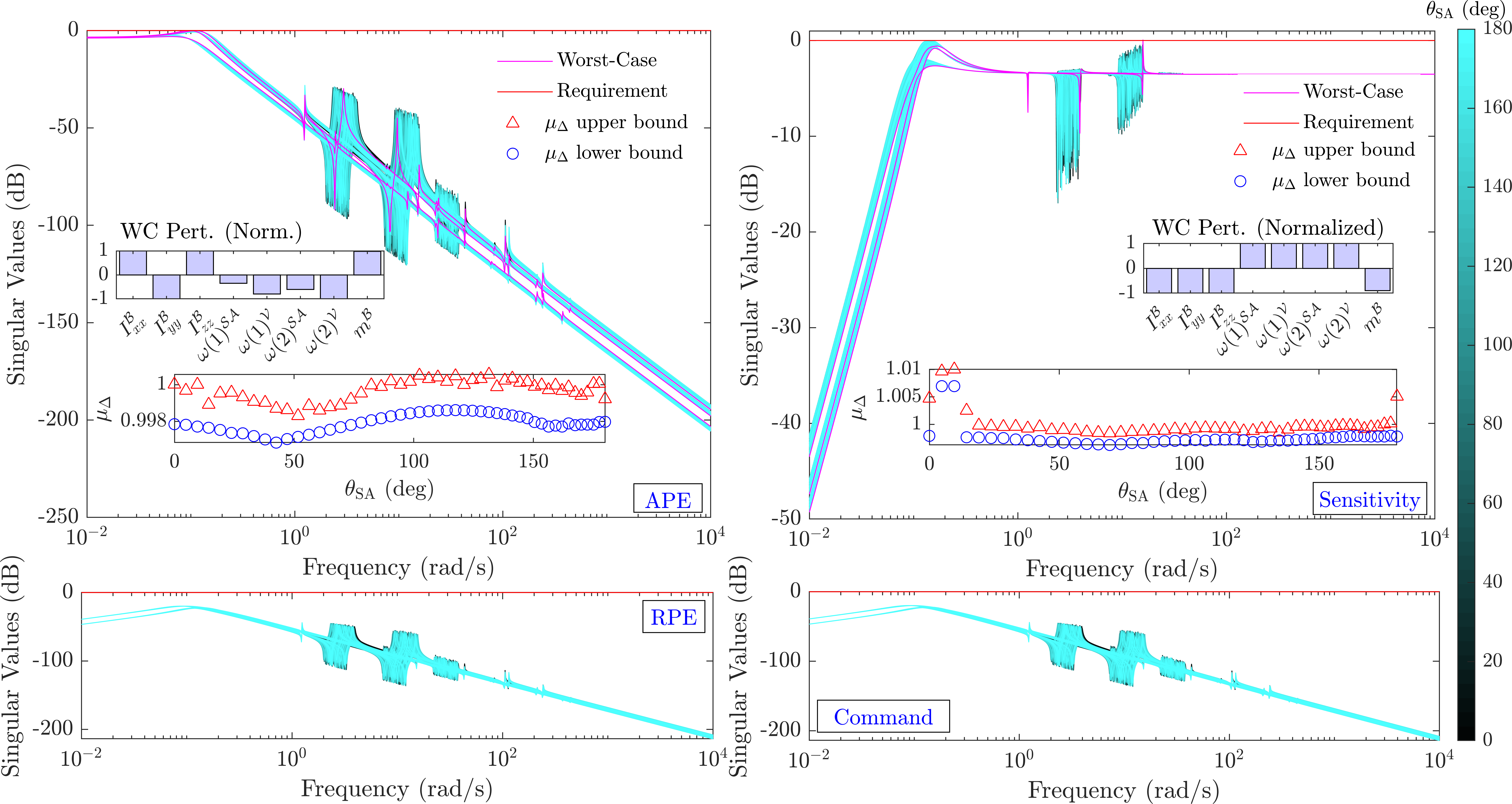}
    \caption{Worst-case analysis of control performance for monolithic optimization}
    \label{fig:ControlPerf_Monolithic}
\end{figure*}

Figure \ref{fig:ControlPerf_Distributed} and \ref{fig:ControlPerf_Monolithic} show the results of this analysis for the distributed and monolithic architectures respectively.
Note that for this validation the optimal parameters obtained with the monolithic approach are used to generate a new dynamical model directly with NASTRAN in order to avoid the use of the model based on the fitted block $\bm\Pi^\mathcal{SC}$.

One can check that the gaps between the upper and the lower bounds of $\mu_\Delta$ are tight by showing that the WC performance is accurately evaluated and very close to the value provided by \textit{systune}. In the same figure the WC combination of parameters corresponding to the overall worst-case $\mu_\Delta$ is provided as well. For both optimization architectures, the largest degradation is obtained for the minimal value of the central body inertia, that can be physically explained with a consequent bigger contribution of the flexible structural parts to the spacecraft dynamics.

\section{Conclusion}
\label{sec:conclusions}

In this work, an end-to-end methodology has been presented for robust design of structure/control co-optimization problems. Both the distributed and a monolithic optimization architectures have been proposed and applied to a scientific ESA spacecraft benchmark. It has been shown how the TITOP modeling approach is well suited to take into account all parametric uncertainties and optimization variable in a unique LFT model, that can be used to obtain a robust optimal solution in a straightforward way. Moreover, the LFT framework is useful for final validation of the achieved control performances.
With both the distributed and monolithic architecture an important reduction of the spacecraft mass is achieved by meeting the mission control performance. This point demonstrated the importance of treating the problem of structure and control optimization in the same framework and avoid the sequential approach traditionally employed in the industrial context.
The limits of both approaches have been also discussed: when using a distributed architecture a longer computational time is the price to pay for including a bigger set of optimization parameters with respect to the monolithic architecture. The LFT complexity drives in fact the maximum number of optimization variables in the monolithic approach.  As seen for the ENVISION benchmark, this parameters' selection brought, as compromise, to a worse control performance, by reaching the same mass reduction as for the distributed optimization. 
It has also been shown that, with the distributed approach, it is possible to introduce exogenous structural constraints (like the launch constraint in this work), which relies on further structural analysis at each iteration. The distributed approach also makes a deeper physical understanding of the achieved results possible by giving access to the evolution of the structural parameters during the optimization iterations. 




\section*{Declarations}
\textbf{Conflict of interest} On behalf of all authors, the corresponding author states that there is no conflict of interest.

\noindent\textbf{Funding} This work was supported by the European Space Agency (ESA) (Contract NO. 4000133632/21/NL/CRS) in the context of the Open Invitations To Tender (ITT) "Control/structure codesign for planetary spacecraft with large flexible appendages" (AO/1-10332/20/NL/CRS).

\noindent\textbf{Replication of results} Upon reasonable request the authors are willing
to share part of the MATLAB code used to generate the results in this paper. Industrial data are not available. 




\bibliography{sn-bibliography}

\begin{thebibliography}{}
\providecommand{\doi}[1]{\url{https://doi.org/#1}}
\bibcommenthead

\bibitem [\protect \citeauthoryear {%
Alavi%
, Dolatabadi%
, Mashhadi%
\BCBL {}\ \BBA {} Noroozinejad~Farsangi%
}{%
Alavi%
\ \protect \BOthers {.}}{%
{\protect \APACyear {2021}}%
}]{%
alavi2021simultaneous}
\APACinsertmetastar {%
alavi2021simultaneous}%
\begin{APACrefauthors}%
Alavi, A.%
, Dolatabadi, M.%
, Mashhadi, J.%
\BCBL {} Noroozinejad~Farsangi, E.%
\end{APACrefauthors}%
\unskip\
\newblock
\APACrefYearMonthDay{2021}{}{}.
\newblock
{\BBOQ}\APACrefatitle {Simultaneous optimization approach for combined
  control--structural design versus the conventional sequential optimization
  method} {Simultaneous optimization approach for combined control--structural
  design versus the conventional sequential optimization method}.{\BBCQ}
\newblock
\APACjournalVolNumPages{Structural and Multidisciplinary
  Optimization}{63}{}{1367--1383}.
\newblock

\newblock

\PrintBackRefs{\CurrentBib}

\bibitem [\protect \citeauthoryear {%
Alazard%
, Finozzi%
\BCBL {}\ \BBA {} Sanfedino%
}{%
Alazard%
\ \protect \BOthers {.}}{%
{\protect \APACyear {2023}}%
}]{%
alazard2023port}
\APACinsertmetastar {%
alazard2023port}%
\begin{APACrefauthors}%
Alazard, D.%
, Finozzi, A.%
\BCBL {} Sanfedino, F.%
\end{APACrefauthors}%
\unskip\
\newblock
\APACrefYearMonthDay{2023}{}{}.
\newblock
{\BBOQ}\APACrefatitle {Port inversions of parametric Two-Input Two-Output Port
  models of flexible substructures} {Port inversions of parametric two-input
  two-output port models of flexible substructures}.{\BBCQ}
\newblock
\APACjournalVolNumPages{Multibody System Dynamics}{57}{3-4}{365--387}.
\newblock

\newblock

\PrintBackRefs{\CurrentBib}

\bibitem [\protect \citeauthoryear {%
Alazard%
, Loquen%
, De~Plinval%
\BCBL {}\ \BBA {} Cumer%
}{%
Alazard%
\ \protect \BOthers {.}}{%
{\protect \APACyear {2013}}%
}]{%
alazard2013avionics}
\APACinsertmetastar {%
alazard2013avionics}%
\begin{APACrefauthors}%
Alazard, D.%
, Loquen, T.%
, De~Plinval, H.%
\BCBL {} Cumer, C.%
\end{APACrefauthors}%
\unskip\
\newblock
\APACrefYearMonthDay{2013}{}{}.
\newblock
{\BBOQ}\APACrefatitle {Avionics/Control co-design for large flexible space
  structures} {Avionics/control co-design for large flexible space
  structures}.{\BBCQ}
\newblock
 \APACrefbtitle {AIAA Guidance, Navigation, and Control (GNC) Conference} {Aiaa
  guidance, navigation, and control (gnc) conference}\ (\BPG~4638).
\PrintBackRefs{\CurrentBib}

\bibitem [\protect \citeauthoryear {%
Alazard%
, Perez%
, Cumer%
\BCBL {}\ \BBA {} Loquen%
}{%
Alazard%
\ \protect \BOthers {.}}{%
{\protect \APACyear {2015}}%
}]{%
alazard2015two}
\APACinsertmetastar {%
alazard2015two}%
\begin{APACrefauthors}%
Alazard, D.%
, Perez, J.A.%
, Cumer, C.%
\BCBL {} Loquen, T.%
\end{APACrefauthors}%
\unskip\
\newblock
\APACrefYearMonthDay{2015}{}{}.
\newblock
{\BBOQ}\APACrefatitle {Two-input two-output port model for mechanical systems}
  {Two-input two-output port model for mechanical systems}.{\BBCQ}
\newblock
 \APACrefbtitle {AIAA Guidance, Navigation, and Control Conference} {Aiaa
  guidance, navigation, and control conference}\ (\BPG~1778).
\PrintBackRefs{\CurrentBib}

\bibitem [\protect \citeauthoryear {%
Alazard%
\ \BBA {} Sanfedino%
}{%
Alazard%
\ \BBA {} Sanfedino%
}{%
{\protect \APACyear {2020}}%
}]{%
sdt}
\APACinsertmetastar {%
sdt}%
\begin{APACrefauthors}%
Alazard, D.%
\BCBT {}\ \BBA {} Sanfedino, F.%
\end{APACrefauthors}%
\unskip\
\newblock
\APACrefYearMonthDay{2020}{}{}.
\newblock
{\BBOQ}\APACrefatitle {Satellite dynamics toolbox for preliminary design phase}
  {Satellite dynamics toolbox for preliminary design phase}.{\BBCQ}
\newblock
\APACjournalVolNumPages{43rd Annual AAS Guidance and Control
  Conf.}{30}{}{1461--1472}.
\newblock

\newblock

\PrintBackRefs{\CurrentBib}

\bibitem [\protect \citeauthoryear {%
Allison%
, Guo%
\BCBL {}\ \BBA {} Han%
}{%
Allison%
\ \protect \BOthers {.}}{%
{\protect \APACyear {2014}}%
}]{%
Allison2014}
\APACinsertmetastar {%
Allison2014}%
\begin{APACrefauthors}%
Allison, J.T.%
, Guo, T.%
\BCBL {} Han, Z.%
\end{APACrefauthors}%
\unskip\
\newblock
\APACrefYearMonthDay{2014}{06}{}.
\newblock
{\BBOQ}\APACrefatitle {{Co-Design of an Active Suspension Using Simultaneous
  Dynamic Optimization}} {{Co-Design of an Active Suspension Using Simultaneous
  Dynamic Optimization}}.{\BBCQ}
\newblock
\APACjournalVolNumPages{Journal of Mechanical Design}{136}{8}{}.
\newblock

\newblock

\PrintBackRefs{\CurrentBib}

\bibitem [\protect \citeauthoryear {%
Apkarian%
, Bompart%
\BCBL {}\ \BBA {} Noll%
}{%
Apkarian%
\ \protect \BOthers {.}}{%
{\protect \APACyear {2007}}%
}]{%
apk2007}
\APACinsertmetastar {%
apk2007}%
\begin{APACrefauthors}%
Apkarian, P.%
, Bompart, V.%
\BCBL {} Noll, D.%
\end{APACrefauthors}%
\unskip\
\newblock
\APACrefYearMonthDay{2007}{}{}.
\newblock
{\BBOQ}\APACrefatitle {Non-smooth structured control design with application to
  PID loop-shaping of a process} {Non-smooth structured control design with
  application to pid loop-shaping of a process}.{\BBCQ}
\newblock
\APACjournalVolNumPages{International Journal of Robust and Nonlinear
  Control}{17}{14}{1320-1342}.
\newblock

\newblock

\newblock
\begin{APACrefDOI} \doi{https://doi.org/10.1002/rnc.1175} \end{APACrefDOI}
\PrintBackRefs{\CurrentBib}

\bibitem [\protect \citeauthoryear {%
Apkarian%
, Dao%
\BCBL {}\ \BBA {} Noll%
}{%
Apkarian%
\ \protect \BOthers {.}}{%
{\protect \APACyear {2015}}%
}]{%
apkarian2015parametric}
\APACinsertmetastar {%
apkarian2015parametric}%
\begin{APACrefauthors}%
Apkarian, P.%
, Dao, M.N.%
\BCBL {} Noll, D.%
\end{APACrefauthors}%
\unskip\
\newblock
\APACrefYearMonthDay{2015}{}{}.
\newblock
{\BBOQ}\APACrefatitle {Parametric robust structured control design} {Parametric
  robust structured control design}.{\BBCQ}
\newblock
\APACjournalVolNumPages{IEEE Trans. on Automatic Control}{60}{7}{1857--1869}.
\newblock

\newblock

\PrintBackRefs{\CurrentBib}

\bibitem [\protect \citeauthoryear {%
Chebbi%
, Dubanchet%
, Gonzalez%
\BCBL {}\ \BBA {} Alazard%
}{%
Chebbi%
\ \protect \BOthers {.}}{%
{\protect \APACyear {2017}}%
}]{%
lin_dyn_flex}
\APACinsertmetastar {%
lin_dyn_flex}%
\begin{APACrefauthors}%
Chebbi, J.%
, Dubanchet, V.%
, Gonzalez, J.A.P.%
\BCBL {} Alazard, D.%
\end{APACrefauthors}%
\unskip\
\newblock
\APACrefYearMonthDay{2017}{}{}.
\newblock
{\BBOQ}\APACrefatitle {Linear dynamics of flexible multibody systems} {Linear
  dynamics of flexible multibody systems}.{\BBCQ}
\newblock
\APACjournalVolNumPages{Multibody System Dynamics}{}{}{}.
\newblock

\newblock

\newblock
\begin{APACrefDOI} \doi{10.1007/s11044-016-9559-y} \end{APACrefDOI}
\PrintBackRefs{\CurrentBib}

\bibitem [\protect \citeauthoryear {%
Chen%
\ \BBA {} Cheng%
}{%
Chen%
\ \BBA {} Cheng%
}{%
{\protect \APACyear {2006}}%
}]{%
chen20063d}
\APACinsertmetastar {%
chen20063d}%
\begin{APACrefauthors}%
Chen, C.Y.%
\BCBT {}\ \BBA {} Cheng, C.C.%
\end{APACrefauthors}%
\unskip\
\newblock
\APACrefYearMonthDay{2006}{}{}.
\newblock
{\BBOQ}\APACrefatitle {3d model based design for control of a mechatronic
  machine tools system} {3d model based design for control of a mechatronic
  machine tools system}.{\BBCQ}
\newblock
 (\BVOL~505, \BPGS\ 967--972).
\newblock
\APACrefnote{Materials science forum}
\PrintBackRefs{\CurrentBib}

\bibitem [\protect \citeauthoryear {%
Chilan%
\ \protect \BOthers {.}}{%
Chilan%
\ \protect \BOthers {.}}{%
{\protect \APACyear {2017}}%
}]{%
Chilan2017}
\APACinsertmetastar {%
Chilan2017}%
\begin{APACrefauthors}%
Chilan, C.M.%
, Herber, D.R.%
, Nakka, Y.K.%
, Chung, S\BHBI J.%
, Allison, J.T.%
, Aldrich, J.B.%
\BCBL {} Alvarez-Salazar, O.S.%
\end{APACrefauthors}%
\unskip\
\newblock
\APACrefYearMonthDay{2017}{}{}.
\newblock
{\BBOQ}\APACrefatitle {Co-Design of Strain-Actuated Solar Arrays for Spacecraft
  Precision Pointing and Jitter Reduction} {Co-design of strain-actuated solar
  arrays for spacecraft precision pointing and jitter reduction}.{\BBCQ}
\newblock
\APACjournalVolNumPages{AIAA Journal}{55}{9}{3180-3195}.
\newblock

\newblock

\PrintBackRefs{\CurrentBib}

\bibitem [\protect \citeauthoryear {%
Dubanchet%
}{%
Dubanchet%
}{%
{\protect \APACyear {2016}}%
}]{%
dubanchet2016modeling}
\APACinsertmetastar {%
dubanchet2016modeling}%
\begin{APACrefauthors}%
Dubanchet, V.%
\end{APACrefauthors}%
\unskip\
\newblock
\APACrefYear{2016}.
\unskip\
\newblock
\APACrefbtitle {Modeling and control of a flexible space robot to capture a
  tumbling debris} {Modeling and control of a flexible space robot to capture a
  tumbling debris}\ \APACtypeAddressSchool {\BUPhD}{}{}.
\PrintBackRefs{\CurrentBib}

\bibitem [\protect \citeauthoryear {%
Falcoz%
\ \protect \BOthers {.}}{%
Falcoz%
\ \protect \BOthers {.}}{%
{\protect \APACyear {2013}}%
}]{%
FALCOZ201313}
\APACinsertmetastar {%
FALCOZ201313}%
\begin{APACrefauthors}%
Falcoz, A.%
, Watt, M.%
, Yu, M.%
, Kron, A.%
, Menon, P.P.%
, Bates, D.%
\BDBL {}Massotti, L.%
\end{APACrefauthors}%
\unskip\
\newblock
\APACrefYearMonthDay{2013}{}{}.
\newblock
{\BBOQ}\APACrefatitle {Integrated Control and Structure design framework for
  spacecraft applied to Biomass satellite.} {Integrated control and structure
  design framework for spacecraft applied to biomass satellite.}{\BBCQ}
\newblock
\APACjournalVolNumPages{IFAC Proceedings Volumes}{46}{19}{13-18}.
\newblock
\APACrefnote{19th IFAC Symposium on Automatic Control in Aerospace}
\newblock

\newblock

\PrintBackRefs{\CurrentBib}

\bibitem [\protect \citeauthoryear {%
Fathy%
, Reyer%
, Papalambros%
\BCBL {}\ \BBA {} Ulsov%
}{%
Fathy%
\ \protect \BOthers {.}}{%
{\protect \APACyear {2001}}%
}]{%
Fathy2001}
\APACinsertmetastar {%
Fathy2001}%
\begin{APACrefauthors}%
Fathy, H.%
, Reyer, J.%
, Papalambros, P.%
\BCBL {} Ulsov, A.%
\end{APACrefauthors}%
\unskip\
\newblock
\APACrefYearMonthDay{2001}{}{}.
\newblock
{\BBOQ}\APACrefatitle {On the coupling between the plant and controller
  optimization problems} {On the coupling between the plant and controller
  optimization problems}.{\BBCQ}
\newblock
 \APACrefbtitle {Proceedings of the 2001 American Control Conference}
  {Proceedings of the 2001 american control conference}\ (\BVOL~3,
  \BPG~1864-1869).
\PrintBackRefs{\CurrentBib}

\bibitem [\protect \citeauthoryear {%
Feng%
, Zhang%
, Tang%
, Yang%
\BCBL {}\ \BBA {} Ge%
}{%
Feng%
\ \protect \BOthers {.}}{%
{\protect \APACyear {2014}}%
}]{%
feng2014control}
\APACinsertmetastar {%
feng2014control}%
\begin{APACrefauthors}%
Feng, Z.%
, Zhang, Q.%
, Tang, Q.%
, Yang, T.%
\BCBL {} Ge, J.%
\end{APACrefauthors}%
\unskip\
\newblock
\APACrefYearMonthDay{2014}{}{}.
\newblock
{\BBOQ}\APACrefatitle {Control-structure integrated multiobjective design for
  flexible spacecraft using MOEA/D} {Control-structure integrated
  multiobjective design for flexible spacecraft using moea/d}.{\BBCQ}
\newblock
\APACjournalVolNumPages{Structural and Multidisciplinary
  Optimization}{50}{}{347--362}.
\newblock

\newblock

\PrintBackRefs{\CurrentBib}

\bibitem [\protect \citeauthoryear {%
Frischknecht%
, Peters%
\BCBL {}\ \BBA {} Papalambros%
}{%
Frischknecht%
\ \protect \BOthers {.}}{%
{\protect \APACyear {2011}}%
}]{%
frischknecht2011pareto}
\APACinsertmetastar {%
frischknecht2011pareto}%
\begin{APACrefauthors}%
Frischknecht, B.D.%
, Peters, D.L.%
\BCBL {} Papalambros, P.Y.%
\end{APACrefauthors}%
\unskip\
\newblock
\APACrefYearMonthDay{2011}{}{}.
\newblock
{\BBOQ}\APACrefatitle {Pareto set analysis: local measures of objective
  coupling in multiobjective design optimization} {Pareto set analysis: local
  measures of objective coupling in multiobjective design optimization}.{\BBCQ}
\newblock
\APACjournalVolNumPages{Structural and Multidisciplinary
  Optimization}{43}{5}{617--630}.
\newblock

\newblock

\PrintBackRefs{\CurrentBib}

\bibitem [\protect \citeauthoryear {%
Gahinet%
\ \BBA {} Apkarian%
}{%
Gahinet%
\ \BBA {} Apkarian%
}{%
{\protect \APACyear {2011}}%
}]{%
GAHINET20111435}
\APACinsertmetastar {%
GAHINET20111435}%
\begin{APACrefauthors}%
Gahinet, P.%
\BCBT {}\ \BBA {} Apkarian, P.%
\end{APACrefauthors}%
\unskip\
\newblock
\APACrefYearMonthDay{2011}{}{}.
\newblock
{\BBOQ}\APACrefatitle {Structured $\mathcal{H}_\infty$ Synthesis in MATLAB}
  {Structured $\mathcal{H}_\infty$ synthesis in matlab}.{\BBCQ}
\newblock
\APACjournalVolNumPages{IFAC Proceedings Volumes}{44}{1}{1435-1440}.
\newblock
\APACrefnote{18th IFAC World Congress}
\newblock

\newblock

\PrintBackRefs{\CurrentBib}

\bibitem [\protect \citeauthoryear {%
Kennedy%
\ \BBA {} Eberhart%
}{%
Kennedy%
\ \BBA {} Eberhart%
}{%
{\protect \APACyear {1995}}%
}]{%
kennedy95}
\APACinsertmetastar {%
kennedy95}%
\begin{APACrefauthors}%
Kennedy, J.%
\BCBT {}\ \BBA {} Eberhart, R.%
\end{APACrefauthors}%
\unskip\
\newblock
\APACrefYearMonthDay{1995}{}{}.
\newblock
{\BBOQ}\APACrefatitle {Particle swarm optimization} {Particle swarm
  optimization}.{\BBCQ}
\newblock
 \APACrefbtitle {Proceedings of ICNN'95 - International Conference on Neural
  Networks} {Proceedings of icnn'95 - international conference on neural
  networks}\ (\BVOL~4, \BPG~1942-1948 vol.4).
\newblock
\begin{APACrefDOI} \doi{10.1109/ICNN.1995.488968} \end{APACrefDOI}
\PrintBackRefs{\CurrentBib}

\bibitem [\protect \citeauthoryear {%
Li%
, Zhang%
\BCBL {}\ \BBA {} Chen%
}{%
Li%
\ \protect \BOthers {.}}{%
{\protect \APACyear {2001}}%
}]{%
Li2001}
\APACinsertmetastar {%
Li2001}%
\begin{APACrefauthors}%
Li, Q.%
, Zhang, W.%
\BCBL {} Chen, L.%
\end{APACrefauthors}%
\unskip\
\newblock
\APACrefYearMonthDay{2001}{}{}.
\newblock
{\BBOQ}\APACrefatitle {Design for control-a concurrent engineering approach for
  mechatronic systems design} {Design for control-a concurrent engineering
  approach for mechatronic systems design}.{\BBCQ}
\newblock
\APACjournalVolNumPages{IEEE/ASME Transactions on Mechatronics}{6}{2}{161-169}.
\newblock

\newblock

\PrintBackRefs{\CurrentBib}

\bibitem [\protect \citeauthoryear {%
Maraniello%
\ \BBA {} Palacios%
}{%
Maraniello%
\ \BBA {} Palacios%
}{%
{\protect \APACyear {2016}}%
}]{%
MARANIELLO20161}
\APACinsertmetastar {%
MARANIELLO20161}%
\begin{APACrefauthors}%
Maraniello, S.%
\BCBT {}\ \BBA {} Palacios, R.%
\end{APACrefauthors}%
\unskip\
\newblock
\APACrefYearMonthDay{2016}{}{}.
\newblock
{\BBOQ}\APACrefatitle {Optimal vibration control and co-design of very flexible
  actuated structures} {Optimal vibration control and co-design of very
  flexible actuated structures}.{\BBCQ}
\newblock
\APACjournalVolNumPages{Journal of Sound and Vibration}{377}{}{1-21}.
\newblock

\newblock

\PrintBackRefs{\CurrentBib}

\bibitem [\protect \citeauthoryear {%
Martins%
\ \BBA {} Lambe%
}{%
Martins%
\ \BBA {} Lambe%
}{%
{\protect \APACyear {2013}}%
}]{%
martins2013multidisciplinary}
\APACinsertmetastar {%
martins2013multidisciplinary}%
\begin{APACrefauthors}%
Martins, J.R.%
\BCBT {}\ \BBA {} Lambe, A.B.%
\end{APACrefauthors}%
\unskip\
\newblock
\APACrefYearMonthDay{2013}{}{}.
\newblock
{\BBOQ}\APACrefatitle {Multidisciplinary design optimization: a survey of
  architectures} {Multidisciplinary design optimization: a survey of
  architectures}.{\BBCQ}
\newblock
\APACjournalVolNumPages{AIAA journal}{51}{9}{2049--2075}.
\newblock

\newblock

\PrintBackRefs{\CurrentBib}

\bibitem [\protect \citeauthoryear {%
Perez%
, Pittet%
, Alazard%
, Loquen%
\BCBL {}\ \BBA {} Cumer%
}{%
Perez%
\ \protect \BOthers {.}}{%
{\protect \APACyear {2015}}%
}]{%
PEREZ2015275}
\APACinsertmetastar {%
PEREZ2015275}%
\begin{APACrefauthors}%
Perez, J.A.%
, Pittet, C.%
, Alazard, D.%
, Loquen, T.%
\BCBL {} Cumer, C.%
\end{APACrefauthors}%
\unskip\
\newblock
\APACrefYearMonthDay{2015}{}{}.
\newblock
{\BBOQ}\APACrefatitle {A Flexible Appendage Model for Use in Integrated
  Control/Structure Spacecraft Design} {A flexible appendage model for use in
  integrated control/structure spacecraft design}.{\BBCQ}
\newblock
\APACjournalVolNumPages{IFAC-PapersOnLine}{48}{9}{275-280}.
\newblock
\APACrefnote{1st IFAC Workshop on Advanced Control and Navigation for
  Autonomous Aerospace Vehicles ACNAAV’15}
\newblock

\newblock

\PrintBackRefs{\CurrentBib}

\bibitem [\protect \citeauthoryear {%
Poussot-Vassal%
\ \BBA {} Roos%
}{%
Poussot-Vassal%
\ \BBA {} Roos%
}{%
{\protect \APACyear {2012}}%
}]{%
poussot2012generation}
\APACinsertmetastar {%
poussot2012generation}%
\begin{APACrefauthors}%
Poussot-Vassal, C.%
\BCBT {}\ \BBA {} Roos, C.%
\end{APACrefauthors}%
\unskip\
\newblock
\APACrefYearMonthDay{2012}{}{}.
\newblock
{\BBOQ}\APACrefatitle {Generation of a reduced-order LPV/LFT model from a set
  of large-scale MIMO LTI flexible aircraft models} {Generation of a
  reduced-order lpv/lft model from a set of large-scale mimo lti flexible
  aircraft models}.{\BBCQ}
\newblock
\APACjournalVolNumPages{Control Engineering Practice}{20}{9}{919--930}.
\newblock

\newblock

\PrintBackRefs{\CurrentBib}

\bibitem [\protect \citeauthoryear {%
Reyer%
, Fathy%
, Papalambros%
\BCBL {}\ \BBA {} Ulsoy%
}{%
Reyer%
\ \protect \BOthers {.}}{%
{\protect \APACyear {2001}}%
}]{%
Reyer2001}
\APACinsertmetastar {%
Reyer2001}%
\begin{APACrefauthors}%
Reyer, J.A.%
, Fathy, H.K.%
, Papalambros, P.Y.%
\BCBL {} Ulsoy, A.G.%
\end{APACrefauthors}%
\unskip\
\newblock
\APACrefYearMonthDay{2001}{09}{}.
\newblock
{\BBOQ}\APACrefatitle {{Comparison of Combined Embodiment Design and Control
  Optimization Strategies Using Optimality Conditions}} {{Comparison of
  Combined Embodiment Design and Control Optimization Strategies Using
  Optimality Conditions}}.{\BBCQ}
\newblock
 (\BPG~1023-1032).
\newblock
\APACrefnote{27th International Design Engineering Technical Conferences and
  Computers and Information in Engineering Conference}
\PrintBackRefs{\CurrentBib}

\bibitem [\protect \citeauthoryear {%
Roos%
, Hardier%
\BCBL {}\ \BBA {} Biannic%
}{%
Roos%
\ \protect \BOthers {.}}{%
{\protect \APACyear {2014}}%
}]{%
roos2014polynomial}
\APACinsertmetastar {%
roos2014polynomial}%
\begin{APACrefauthors}%
Roos, C.%
, Hardier, G.%
\BCBL {} Biannic, J\BHBI M.%
\end{APACrefauthors}%
\unskip\
\newblock
\APACrefYearMonthDay{2014}{}{}.
\newblock
{\BBOQ}\APACrefatitle {Polynomial and rational approximation with the APRICOT
  library of the SMAC toolbox} {Polynomial and rational approximation with the
  apricot library of the smac toolbox}.{\BBCQ}
\newblock
 \APACrefbtitle {2014 IEEE Conference on Control Applications (CCA)} {2014 ieee
  conference on control applications (cca)}\ (\BPGS\ 1473--1478).
\PrintBackRefs{\CurrentBib}

\bibitem [\protect \citeauthoryear {%
Sanfedino%
}{%
Sanfedino%
}{%
{\protect \APACyear {2019}}%
}]{%
sanfedino2019experimental}
\APACinsertmetastar {%
sanfedino2019experimental}%
\begin{APACrefauthors}%
Sanfedino, F.%
\end{APACrefauthors}%
\unskip\
\newblock
\APACrefYear{2019}.
\unskip\
\newblock
\APACrefbtitle {Experimental validation of a high accuracy pointing system}
  {Experimental validation of a high accuracy pointing system}\
  \APACtypeAddressSchool {\BUPhD}{}{}.
\unskip\
\newblock
\APACaddressSchool {}{ISAE-SUPAERO, Toulouse (France)}.
\PrintBackRefs{\CurrentBib}

\bibitem [\protect \citeauthoryear {%
Sanfedino%
, Alazard%
, Pommier-Budinger%
, Falcoz%
\BCBL {}\ \BBA {} Boquet%
}{%
Sanfedino%
\ \protect \BOthers {.}}{%
{\protect \APACyear {2018}}%
}]{%
sanfedino2018finite}
\APACinsertmetastar {%
sanfedino2018finite}%
\begin{APACrefauthors}%
Sanfedino, F.%
, Alazard, D.%
, Pommier-Budinger, V.%
, Falcoz, A.%
\BCBL {} Boquet, F.%
\end{APACrefauthors}%
\unskip\
\newblock
\APACrefYearMonthDay{2018}{}{}.
\newblock
{\BBOQ}\APACrefatitle {Finite element based n-port model for preliminary design
  of multibody systems} {Finite element based n-port model for preliminary
  design of multibody systems}.{\BBCQ}
\newblock
\APACjournalVolNumPages{Journal of Sound and Vibration}{415}{}{128--146}.
\newblock

\newblock

\PrintBackRefs{\CurrentBib}

\bibitem [\protect \citeauthoryear {%
Sanfedino%
, Alazard%
, Preda%
\BCBL {}\ \BBA {} Oddenino%
}{%
Sanfedino%
\ \protect \BOthers {.}}{%
{\protect \APACyear {2022}}%
}]{%
sanfedino2022integrated}
\APACinsertmetastar {%
sanfedino2022integrated}%
\begin{APACrefauthors}%
Sanfedino, F.%
, Alazard, D.%
, Preda, V.%
\BCBL {} Oddenino, D.%
\end{APACrefauthors}%
\unskip\
\newblock
\APACrefYearMonthDay{2022}{}{}.
\newblock
{\BBOQ}\APACrefatitle {Integrated modeling of microvibrations induced by Solar
  Array Drive Mechanism for worst-case end-to-end analysis and robust
  disturbance estimation} {Integrated modeling of microvibrations induced by
  solar array drive mechanism for worst-case end-to-end analysis and robust
  disturbance estimation}.{\BBCQ}
\newblock
\APACjournalVolNumPages{Mechanical Systems and Signal
  Processing}{163}{}{108168}.
\newblock

\newblock

\PrintBackRefs{\CurrentBib}

\bibitem [\protect \citeauthoryear {%
Toglia%
\ \protect \BOthers {.}}{%
Toglia%
\ \protect \BOthers {.}}{%
{\protect \APACyear {2013}}%
}]{%
toglia2013optimal}
\APACinsertmetastar {%
toglia2013optimal}%
\begin{APACrefauthors}%
Toglia, C.%
, Pavia, P.%
, Campolo, G.%
, Alazard, D.%
, Loquen, T.%
, de Plinval, H.%
\BDBL {}Massotti, L.%
\end{APACrefauthors}%
\unskip\
\newblock
\APACrefYearMonthDay{2013}{}{}.
\newblock
{\BBOQ}\APACrefatitle {Optimal co-design for earth observation satellites with
  flexible appendages} {Optimal co-design for earth observation satellites with
  flexible appendages}.{\BBCQ}
\newblock
 \APACrefbtitle {AIAA Guidance, Navigation, and Control (GNC) Conference} {Aiaa
  guidance, navigation, and control (gnc) conference}\ (\BPG~4640).
\PrintBackRefs{\CurrentBib}

\bibitem [\protect \citeauthoryear {%
Zhao%
, Chen%
\BCBL {}\ \BBA {} Gu%
}{%
Zhao%
\ \protect \BOthers {.}}{%
{\protect \APACyear {2009}}%
}]{%
zhao2009control}
\APACinsertmetastar {%
zhao2009control}%
\begin{APACrefauthors}%
Zhao, G.%
, Chen, B.%
\BCBL {} Gu, Y.%
\end{APACrefauthors}%
\unskip\
\newblock
\APACrefYearMonthDay{2009}{}{}.
\newblock
{\BBOQ}\APACrefatitle {Control--structural design optimization for vibration of
  piezoelectric intelligent truss structures} {Control--structural design
  optimization for vibration of piezoelectric intelligent truss
  structures}.{\BBCQ}
\newblock
\APACjournalVolNumPages{Structural and Multidisciplinary
  Optimization}{37}{}{509--519}.
\newblock

\newblock

\PrintBackRefs{\CurrentBib}

\bibitem [\protect \citeauthoryear {%
Zheng%
, Zhang%
\BCBL {}\ \BBA {} Zhao%
}{%
Zheng%
\ \protect \BOthers {.}}{%
{\protect \APACyear {2021}}%
}]{%
zheng2021integrated}
\APACinsertmetastar {%
zheng2021integrated}%
\begin{APACrefauthors}%
Zheng, H.%
, Zhang, S.%
\BCBL {} Zhao, G.%
\end{APACrefauthors}%
\unskip\
\newblock
\APACrefYearMonthDay{2021}{}{}.
\newblock
{\BBOQ}\APACrefatitle {Integrated design optimization of actuator layout and
  structural ply parameters for the dynamic shape control of piezoelectric
  laminated curved shell structures} {Integrated design optimization of
  actuator layout and structural ply parameters for the dynamic shape control
  of piezoelectric laminated curved shell structures}.{\BBCQ}
\newblock
\APACjournalVolNumPages{Structural and Multidisciplinary
  Optimization}{63}{}{2375--2398}.
\newblock

\newblock

\PrintBackRefs{\CurrentBib}

\bibitem [\protect \citeauthoryear {%
Zhou%
\ \BBA {} Doyle%
}{%
Zhou%
\ \BBA {} Doyle%
}{%
{\protect \APACyear {1998}}%
}]{%
zhou1998essentials}
\APACinsertmetastar {%
zhou1998essentials}%
\begin{APACrefauthors}%
Zhou, K.%
\BCBT {}\ \BBA {} Doyle, J.C.%
\end{APACrefauthors}%
\unskip\
\newblock
\APACrefYear{1998}.
\newblock
\APACrefbtitle {Essentials of robust control} {Essentials of robust control}.
\newblock
\APACaddressPublisher{}{Prentice hall Upper Saddle River, NJ}.
\PrintBackRefs{\CurrentBib}

\end{thebibliography}


\end{document}